\documentclass[reprint,pra,amsmath,amssymb,superscriptaddress,floatfix,aps]{revtex4-2}
\usepackage[utf8]{inputenc}   
\usepackage{amsmath,amssymb,bm}
\usepackage{dsfont}
\usepackage{comment}
\usepackage{color}
\usepackage{float}
\usepackage{appendix}
\usepackage{soul}
\usepackage{hyperref}
\usepackage{braket}

\usepackage[normalem]{ulem}
\hypersetup{colorlinks,
	linkcolor=blue,%
	citecolor=blue,%
	urlcolor=blue}
\usepackage[T1]{fontenc}

\newcommand{\me}[1]{\textcolor{black}{#1}}
\usepackage{graphicx}
\usepackage{xcolor}
\usepackage{bbding}
\usepackage{tikz}
\usepackage{xcolor}
\usepackage{pifont}
\usepackage[utf8]{inputenc}

\allowdisplaybreaks

\begin{document}
\title{Emergence of giant vortices under nonlinear rotation with attractive interactions in a toroidal condensate}

\author{Rony Boral}
\affiliation{Department of Physics, Indian Institute of Technology, Guwahati 781039, Assam, India} 

\author{Swarup K. Sarkar}
\affiliation{Department of Physics, Indian Institute of Technology, Guwahati 781039, Assam, India} 
\affiliation{Centre for computational and data sciences, Indian Institute of Technology Kharagpur, Kharagpur, West Bengal 721302, India}

\author{Matthew Edmonds}
\affiliation{International Institute for Sustainability with Knotted Chiral Meta Matter (WPI-SKCM$^2$), Hiroshima University, 1-3-1 Kagamiyama, Higashi-Hiroshima, Hiroshima 739-8531, Japan}
\affiliation{Department of Physics and Research and Education Center for Natural Sciences, Keio University, Hiyoshi 4-1-1, Yokohama, Kanagawa 223-8521, Japan} 

\author{Paulsamy Muruganandam}
\affiliation{Department of Physics, Bharathidasan University, Tiruchirappalli 620024, Tamilnadu, India}
\affiliation{Department of Medical Physics, Bharathidasan University, Tiruchirappalli 620024, Tamilnadu, India}

\author{Pankaj Kumar Mishra}
\affiliation{Department of Physics, Indian Institute of Technology, Guwahati 781039, Assam, India}

\date{\today}

\begin{abstract}
We numerically investigate the effects of a density-dependent gauge potential which induces nonlinear rotation on a Bose-Einstein condensate confined in a toroidal trapping geometry. By focusing on the resulting vortex lattice configurations, we demonstrate that increasing the strength of the nonlinear rotation leads to a structural transition from a ring-shaped vortex lattice to a giant vortex state. The quantum circulation associated with the giant vortex is found to be highly sensitive to the strength of the nonlinear rotation and the giant vortex appears in the regime of negative chemical potential. Additionally, we identify the parameter regime in which the Thomas-Fermi density profile remains valid by mapping the solution space based on the strength of the nonlinear rotation and the radius of the confining potential. Based on the Bogoliubov de Gennes analysis, we investigate the impact of nonlinear rotation on the collective excitation spectrum. Our results reveal a violation of the Kohn theorem, accompanied by changes in the breathing mode frequency that indicate radial deformation of the condensate. Our findings are further substantiated through a comprehensive hydrodynamic analysis. Finally, we analyze the stability of multiply quantized vortices under nonlinear rotation. Our findings indicate that while nonlinear rotation can enhance the global stability of these states, it does not necessarily ensure their local stability.
\end{abstract}

\maketitle
\section{Introduction}
The study of topological excitations in quantum fluids has become a cornerstone of modern condensed matter physics, owing to its profound implications across a wide range of physical systems, including superfluid helium~\cite{Donnelly:1991}, liquid crystals~\cite{Malomed:2005}, quantum magnets~\cite{Zapf:2014}, and topological matter~\cite{Sonin:1987}. These excitations—often realized as vortices, solitons, or other defect structures—encode the nontrivial topology of the underlying order parameter and play a decisive role in determining macroscopic properties, collective dynamics, and phase transitions in many-body systems~\cite{Anderson:2007, Leggatt:2006}. Understanding their formation, stability, and dynamics is therefore essential for unraveling the complex behavior of quantum fluids both in and out of equilibrium.

Among the various experimental platforms available, rotating Bose-Einstein condensates (BECs) stand out as an exceptionally clean and versatile system for investigating topological phenomena. The controlled generation of quantized vortices and vortex lattices in BECs offers a direct and highly tunable realization of topological defects in a quantum fluid~\cite{Fetter:2001, estrada2025many}. A particularly rich setting arises in toroidally trapped BECs, where the multiply connected geometry supports persistent currents and long-lived superflow~\cite{Gupta:2005, Olson:2007, Henderson:2009}. These systems have attracted significant theoretical and experimental interest, leading to investigations of persistent currents~\cite{Javanainen:1998, Dubessy:2012}, ring dark solitons~\cite{Toikka:2012, Toikka:2013}, weak links and atomtronic analogs of Josephson junctions~\cite{Ramanathan:2011, Wright:2013}, as well as spontaneous symmetry breaking and metastability~\cite{Oliinyk:2019}. Under rotation the annular geometry gives rise to an especially diverse array of vortex configurations, ranging from regular vortex lattices to giant vortices, with their stability governed by the interplay between angular velocity, interaction strength, and trap geometry~\cite{Fetter:2005, Zaremba:2006, Karabulut:2013}.

Collective excitations play a central role in determining the stability and dynamical response of vortex lattices~\cite{Boral:2025}. In BECs, extensive theoretical and experimental efforts have characterized the excitation spectrum across different regimes~\cite{Ryu:2007, Jackson:2006, Cozzini:2005, Cozzini:2006, Woo:2012, Sun:2018, Padavic:2017}. Analytical results in the rapid-rotation limit were developed in Refs.~\cite{Cozzini:2005, Cozzini:2006}, while the dependence of normal modes on radial and azimuthal quantum numbers has been systematically explored~\cite{Ryu:2007, Ramanathan:2011, Jackson:2006}. In annular condensates, surface excitations exhibit distinct features such as inner-surface modes which are linked to vortex-dipole formation, unlike harmonic traps where vortex monopoles typically emerge~\cite{Woo:2012}. The evolution of excitation spectra from filled to hollow geometries, including gravitational effects, has also been examined~\cite{Sun:2018, Padavic:2017}.


Multicomponent and spinor BECs further enrich this picture, exhibiting modified excitation spectra governed by rotation, interactions, and miscibility~\cite{Shimodira:2010, Wu:2015, Lundh:2013, Yakimenko:2013, Kunimi:2014}. In parallel, the realization of artificial gauge potentials in ultracold gases has opened new avenues for engineering novel quantum states~\cite{Dalibard:2011, Goldman:2014}. Of particular interest are density-dependent (quasi-dynamical) gauge potentials, which introduce nonlinear feedback between the condensate density and effective gauge potential~\cite{Edmonds:2013simulating, Keilmann:2011statistically, Greschner:2015density, Jamotte:2022strain}. These have been explored in both continuum and lattice systems, with experimental demonstrations in bosonic, fermionic, and Rydberg platforms~\cite{Clark:2018, Gorg:2019, Lienhard:2020}, and have led to the experimental observation of domain walls and chiral solitons~\cite{Yao:2022, Frolian:2022, Chrisholm:2022, Iacovelli:2026}. 

The presence of density-dependent gauge potentials significantly alters condensate dynamics, giving rise to nonlinear rotation, irregular and chaotic behavior, and unconventional vortex states~\cite{Edmonds:2015, Zheng:2015, Chen:2022, Butera:2015, Butera:2016, Butera:2017, Correggi:2019}. These effects include the emergence of non-Abrikosov vortex lattices, ring-like vortex configurations~\cite{Edmonds:2020}, modified critical rotation thresholds~\cite{Bhat:2023}, and ghost vortices~\cite{Bhat:2024}, highlighting the rich interplay between topology, interactions, with this form of synthetic gauge potential.

Although several studies have investigated the structural transformation and collective excitation spectrum of vortex lattices under a density-dependent gauge potential in harmonic traps, the influence of such a gauge potential within a toroidal geometry remains unexplored. This represents an intriguing combination to study, as the interplay of the geometry and the toroidal confinement with the induced topology from the artificial gauge potential is anticipated to support unique superfluid phenomenology.

In this work, we provide a comprehensive analysis of the structural transformation of vortex lattices induced by a density-dependent gauge potential in a toroidal trap. We derive the corresponding Thomas-Fermi density profile and identify the parameter regime that supports physically admissible solutions. We also investigate the effect of the density-dependent gauge potential on the size of the giant vortex, which is formed utilizing different techniques such as absorption and repulsive methods. To examine the collective dynamics, we compute the excitation spectrum by solving the Bogoliubov de Gennes (BdG) equations, and further interpret the results within the hydrodynamic framework. Our results reveal that the nonlinear rotation generated by the density-dependent gauge potential induces an energy splitting between excitation modes with angular quantum numbers $l$ and $-l$. Furthermore, we study the effect of the density-dependent gauge potential on the stability multiply quantized vortices.

The paper is organized as follows. In Sec. \ref{sec:meanfieldmodel}, we introduce the mean-field Gross-Pitaevskii model incorporating the nonlinear rotation and describe the numerical methods used. Section \ref{sec:groundstate_TF} presents an analysis of the ground-state properties and the corresponding Thomas-Fermi density profile. In Sec. \ref{sec:absorb_repulse}, we examine how the nonlinear rotation affects the size of the giant vortex generated by various techniques, including absorption and repulsive pinning. Section \ref{sec:collective} explores the influence of the nonlinear rotation on the collective excitation spectrum, while Section \ref{sec:stab_mqv} discusses the stability of multiply quantized vortices in its presence. Finally, in Sec. \ref{sec:conclusion}, we summarize our findings.

\section{Mean-field model}
\label{sec:meanfieldmodel}
The Hamiltonian of a weakly interacting Bose-Einstein condensate, consisting of $N$ two-level atoms coupled through a coherent light-matter interaction, can be written using the rotating wave approximation as~\cite{Dalibard:2011}:
\begin{align}\label{eq:ham}
		\hat{\mathcal{H}} = \left(\frac{\mathbf{\hat{p}}^2}{2m}+ V(\mathbf{r})\right) \otimes \mathds{1} + \hat{\mathcal{H}}_{\text{int}} + \hat{\mathcal{U}}_{\text{MF}},
\end{align}
where $\hat{\mathcal{U}}_{\text{MF}}$ is the light matter interaction which is defined as ~\cite{Goldman:2014} 
\begin{align}\label{eq:lm}
		\hat{\mathcal{U}}_{\text{MF}} = \frac{\hbar \Omega_{r}}{2}
		\begin{pmatrix}
			\cos\theta\left(\mathbf{r}\right)& e^{-i\phi\left(\mathbf{r}\right)}\sin \theta\left(\mathbf{r}\right) \\
			e^{i\phi\left(\mathbf{r}\right)}\sin \theta\left(\mathbf{r}\right) & -\cos \theta\left(\mathbf{r}\right)
		\end{pmatrix}.
\end{align}
Here $\Omega_{r}$ represents the Rabi frequency, $\theta\left(\mathbf{r}\right)$ denotes the mixing angle, and $\phi\left(\mathbf{r}\right)$ indicates the phase of the incident laser beam. 
The momentum operator is given by $\mathbf{\hat{p}}$, and the trapping potential is defined as  $V(\mathbf{r}) = m(\omega^2_{x}x^2 + \omega^2_{y}y^2 + \omega^2_{z}z^2)/2$, while $\mathds{1}$ represents the $2\times 2$ identity matrix in the dressed basis. The mean-field interactions are described by $\hat{\mathcal{H}}_{\text{int}} = (1/2)\text{diag}[\Delta_{1},\Delta_{2}]$, where $\Delta_{i}=g_{ii}\vert \Psi_{i}\vert ^2 + g_{ij}\vert \Psi_{j}\vert ^2$ and $g_{ij} = 4\pi \hbar^2 a_{ij}/m$ with $a_{ij}$ being the s-wave scattering lengths for collisions between atoms in internal states $i$ and $j$ ($i, j = 1, 2$)~\cite{Goldman:2014}. 
To construct the density-dependent gauge theory, we begin with the assumption that the mean field interaction is much weaker than the light-matter coupling. Based on this assumption, we can determine the dressed states using first-order perturbation theory~\cite{Edmonds:2013simulating}.
\begin{align}
		\ket{\Psi_{\pm}} = \ket{\pm} \pm \frac{\Delta_{d}}{\hbar \Omega_{r}}\ket{\mp}
\end{align}
with $\ket{\pm}$ representing the unperturbed eigenstates of $\hat{\mathcal{U}}_{\text{MF}}$, and $\Delta_{d} = \text{sin}\left(\theta/2\right) 
\text{cos}\left(\theta/2\right) \left(\Delta_{1}-\Delta_{2}\right)/2$
denotes the mean-field detuning. 
The effective Hamiltonian is obtained by projecting the atomic motion onto the dressed state basis ${ \ket{\Psi_{\pm}} }$ and can be expressed as
\begin{align}\label{eq:ham+}
		\mathit{\hat{H}}_{+} = \frac{\left(\mathbf{\hat{p}}-\mathbf{A}_{+}\right)^2}{2m} + W_{+} + \frac{\hbar \Omega_{r}}{2} + \Delta_{+} + V (\mathbf{r}),
\end{align}
where $\mathbf{A}_{+} = \mathrm i \hbar\braket{\Psi_{+} \vert \nabla \Psi_{+}}$, and $W_{+} = (\hbar^{2}/2m) \times \vert \braket{\Psi_{+} \vert \nabla \Psi_{-}} \vert^{2} $ are the vector and scalar potentials, respectively \me{while} $\Delta_{+} = \big(\Delta_{1} \cos^{2}(\theta/2) + \Delta_{2} \sin^{2}(\theta/2)\big)/2$ is the dressed mean-field detuning. By invoking the adiabatic approximation and projecting onto the dressed-state $\ket{\Psi_{+}}$, the generalized Gross-Pitaevskii equation in three dimensions can be derived following the procedure explained in \cite{Butera:2017}.
\begin{align}\label{eq:gpe3d}
    \mathrm i \hbar \frac{\partial \Psi}{\partial t} =  \bigg[-\frac{\hbar^2}{2m}\nabla^2 + V(\mathbf{r}) -\Omega_{n}(\mathbf{r},t) \hat{L}_{z} + \text{g}_\text{eff} \vert \Psi\vert ^2 \bigg] \Psi.
\end{align}
In deriving Eq.~\eqref{eq:gpe3d}, we introduce a spatially dependent Rabi coupling of the form $\Omega_{\mathrm{r}} = \kappa_{0} r$, where $r$ denotes the radial distance from the trap center and $\kappa_{0}$ is a constant. The laser field generating this coupling carries an azimuthal phase profile $\phi = l\,\varphi$, where $l$ is the orbital angular momentum quantum number of the light and $\varphi$ denotes the polar angle. The density-dependent rotation \me{is} represented as $\Omega_{n}(\mathbf{r},t)$ is defined in the following way~\cite{Butera:2015}:
\begin{align}\label{eq:om_rho}
		\Omega_{n}(\mathbf{r},t) = \Omega + \mathit{C} n(\mathbf{r}, t),
\end{align} 
here $n(\mathbf{r}, t) = \vert \Psi(\mathbf{r}, t) \vert^2$ represents the condensate density while $\Omega = (\hbar l)/(4m)(\kappa_0/\Delta)^2$ illustrates the controllable rigid-body rotation strength via light-matter interaction. Moreover, the nonlinear rotation strength can be expressed as  $C = l(g_{11} - g_{22})\kappa_{0}^2/(2m\Delta^2)$.

It should be noted that nonlinear rotation \me{could} be experimentally incorporated into the condensate using a laser that carries orbital angular momentum with $l = 1$. This approach is similar in spirit to the creation of spin-angular-momentum-coupled condensates using Laguerre-Gaussian laser beams~\cite{Chen:2018spin, Chen:2018rotating, Zhang:2019ground}. To obtain the quasi-two-dimensional Gross-Pitaevskii (GP) equation, we factorize the condensate wave function as $\Psi(\mathbf{r},t)=\psi(x,y,t)\,\exp\!\left(-z^{2}/2\sigma_{z}^{2}\right)/\left(\pi \sigma_{z}^{2}\right)^{1/4}$, and assume that the condensate is confined in a highly anisotropic harmonic trap such that $\omega_{z} \gg \omega_{x,y}$.
Under this condition, the axial dynamics are frozen into the ground state of the harmonic oscillator. By integrating out the axial degree of freedom, we arrive at the following generalized two-dimensional GP equation:
\begin{align}\label{eq:model}
		\mathrm i \hbar \frac{\partial \psi}{\partial t} =
		\bigg[-\frac{\hbar^2}{2m}\nabla^2 +
		V(x,y) -\Omega_{n} \hat{L}_{z} 
		+ \text{g} \vert \psi\vert ^2 \bigg] \psi.
\end{align}
	
For notational simplicity in what follows we define $\Omega_n({\bf r},t) \equiv \Omega_n,$ and $\psi(x,y,t) \equiv \psi$. The nonlinear coefficients appearing in Eq.~\eqref{eq:model} are \me{rescaled to} $\text{g} \rightarrow \text{g}/\sqrt{2\pi}\sigma_{z}$ and $\mathit{C}\rightarrow\mathit{C}/\sqrt{2\pi}\sigma_z$.
Here, $n(x,y,t) = |\psi(x,y,t)|^2$ denotes the quasi-two-dimensional density of the condensate.
Introducing spatiotemporal scaling, we express time, length, and the order parameter in units of the transverse trapping frequency and oscillator length, 
\begin{align}
t \rightarrow \frac{t}{\omega_{\perp}}, (x,y) \rightarrow a_{\perp} (x,y), \hspace{0.25cm}\text{and}\hspace{0.25cm} \psi\rightarrow\frac{\sqrt{N}}{a_{\perp}}\psi 
\end{align}
where $a_{\perp} = \sqrt{\hbar/m\omega_\perp}$ is the harmonic oscillator length scale. With these rescalings, Eq.~\eqref{eq:model} can be cast into the following dimensionless form:
\begin{align}\label{eq:dmless_eq}
		\mathrm i\frac{\partial\psi}{\partial t} =
		\bigg[-\frac{\nabla^2}{2} +
		V(x,y) -\Omega_{n} \hat{L}_{z} + \text{g} \vert \psi\vert ^2 \bigg]\psi.
\end{align} 
In Eq.~\eqref{eq:dmless_eq}, the scaled parameters are defined as follows: $\text{g} \rightarrow \text{g} N m/\hbar^2$ and $ \Omega_{n}(x,y,t) = \Omega/\omega_{\perp} + \tilde{\mathit{C}} n(x,y,t) $ where $ \tilde{\mathit{C}} = \mathit{C}N /m\hbar\sqrt{2\pi}\sigma_{z}$. The toroidal trap potential is given by $V(x, y) = \frac{1}{2}\left(r - r_0\right)^2$, where $r = \sqrt{x^2 + y^2}$ represents the radial coordinate in the $xy$-plane, and $ r_0$ denotes the toroidal radius at which the potential attains its minimum.
\begin{figure*}[!htp]
\includegraphics[width=0.99\linewidth]{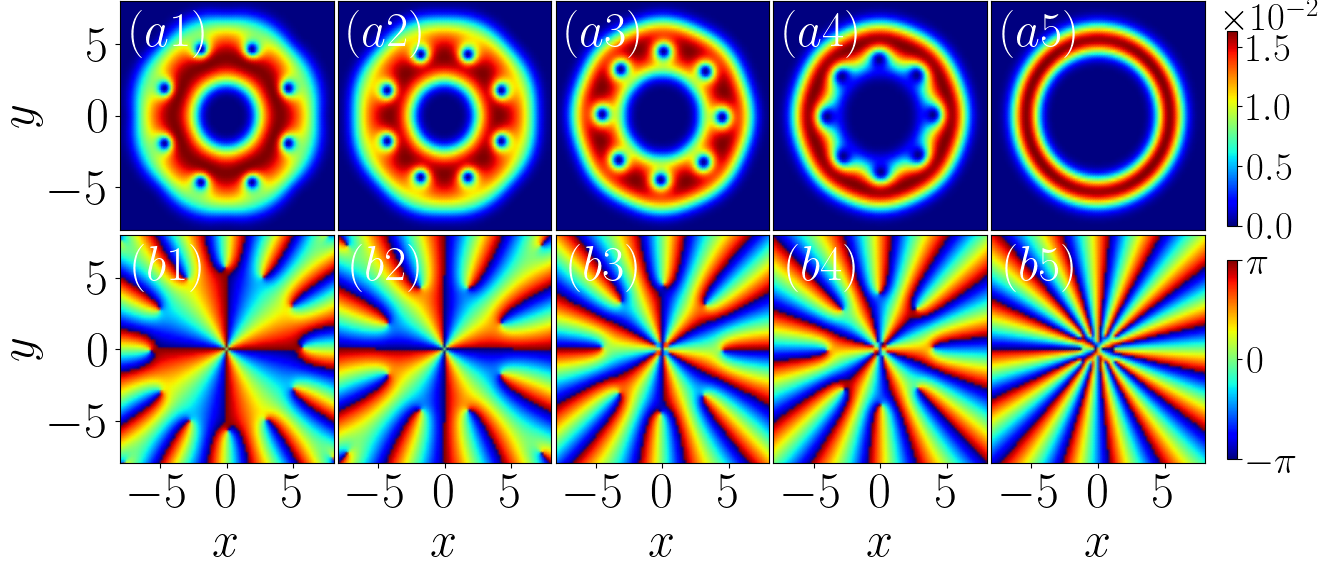}
\caption{Pseudo color representations of the condensate density for $\tilde{C}=-40,-20,0,6,$ and $12$ [(a1)–(a5)] at $\Omega=0.45$ and $\rm g=250$. Panels (b1)–(b5) display the corresponding phase profiles of the densities shown in (a1)–(a5).}
\label{fig:density}
\end{figure*}
    

\section{Ground State and Thomas-Fermi Analysis}
\label{sec:groundstate_TF}
 
We start our analysis by examining the impact of the nonlinear strength $\tilde{C}$ on the ground state of the vortex lattice structure trapped in a toroidal potential. In Fig.~\ref{fig:density}, we present the condensate density and phase profile for different values of $\tilde{C}$ at $\Omega = 0.45$ and $\rm g = 250$, with the radius of the toroidal trap potential fixed at $r_0 = 4$.
At $\tilde{C}=0$, we observe that the condensate shows a ring-shaped vortex lattice structure, featuring a central hole with a total quantum circulation of six, surrounded by a regular array of vortices distributed within the annular region~see [\ref{fig:density}(a3)]. As $\tilde{C}$ continues to increase on the positive side i.e $\tilde{C} > 0$, vortices that are initially situated in the annulus, begin to move towards the center~[\ref{fig:density}(a4)]. This movement leads to an increase in both the size of the hole and its circulation. With a further increase in $\tilde{C}$, the surrounding vortices are localized to the central region, resulting in the formation of a giant vortex~[\ref{fig:density}(a5)].
Conversely, when $\tilde{C}$ is negative, the behavior of the condensate contrasts with that observed for positive $\tilde{C}$. In this scenario, both the central hole region and quantum circulation gradually decrease. Meanwhile, the vortices in the annular region move away from the center, and for sufficiently large negative values of $\tilde{C}$, the vortices become arranged along the periphery of the condensate~[see Figs.~\ref{fig:density}(a1-a2)].
Figure \ref{fig:density}(b1–b5) presents the phase profiles corresponding to the density illustrated in \ref{fig:density}(a1)-(a5). The phase profiles reveal that the multiply quantized central vortex circulation decreases when $\tilde{C}<0$, and increases for $\tilde{C}>0$. For instance, at  $\tilde{C}=0$, the circulation is $6$, which falls to four for $\tilde{C}=-40$, and rises to $15$ for $\tilde{C}=12$.

After obtaining the ground state solution using \me{Eq.~\eqref{eq:dmless_eq}}, we compute the Thomas-Fermi density distribution derived from the energy functional associated with \me{Eq.~\ref{eq:dmless_eq}} for further analysis. In the hydrodynamic picture, using the Madelung transformation $\psi=\sqrt{n}\exp(i\nu S)$, where $n$, $S$ and $\nu$ denote the density of the condensate, phase and quantum circulation, respectively, the functional energy can be expressed as~\cite{Edmonds:2020,Aftalion:2010}
\begin{align}\notag
E_{\rm T}=\int d^2r\ n\bigg[&\frac{1}{2}\frac{|\nabla n|^2}{4n^2}+\frac{\nu^2}{2r^2}+\frac{1}{2}{\bf v}^2-\frac{\mathcal{A}^2}{2}+\frac{i\hbar}{2n}\nabla n\cdot\mathcal{A}\\ \label{eqn:tfen} &+\frac{g}{2}n+V(r)-\mu\bigg]
\end{align}
where ${\bf v}=\big(\hbar\nu\nabla S-\mathcal{A}\big)/m$ defines the hydrodynamic kinetic velocity, while $\mathcal{A}=m\boldsymbol{\Omega}\times{\bf r}$ and $\boldsymbol{\Omega}=\hat{\bf e}_z(\Omega+Cn/2)$. In the Thomas-Fermi limit the quantum pressure term appearing in Eq.~\eqref{eqn:tfen} can be dropped, then by minimizing Eq.~\eqref{eqn:tfen} first with respect to the phase $S$ gives the superfluid velocity ${\bf v}_{\bf sf}={\boldsymbol\Omega}\times{\bf r}/\nu$. Combing ${\bf v}_{\rm sf}$ with Eq.~\eqref{eqn:tfen} and simplifying results in the expression
\begin{align}
\mathrm{E_T}=\int dr n \left[\frac{\nu^{2}}{2r^2}{-}\frac{\mathcal{A}^2}{2}{+}\frac{i}{2n}\nabla n\cdot \mathcal{A}{+}\frac{\mathrm{g} n}{2}{+}V(r){-}\mu\right]
\label{eq:energy}
\end{align}
Here, $V(r)=\frac{1}{2}(r-r_0)^{2}$ is the toroidal trap potential and the term proportional to $\nu^{2}/2r^{2}$ represents the centrifugal contribution associated with the quantized circulation in the central hole of the condensate. To obtain the Thomas-Fermi distribution, we minimize Eq.~\ref{eq:energy} with respect to the the condensate density, which leads to the following expression:

\begin{align}
V(r)-\frac{1}{2}r^2\left(\Omega^2+2\Omega \tilde{C} n+\frac{3}{4} \tilde{C}^2n^2-\frac{\nu^{2}}{r^4}\right)+\mathrm{g}n=\mu
\label{eq:tf_den}
\end{align}

By solving the accompanying quadratic equation for the density $n$, ~(Eq.~\ref{eq:tf_den}) we can obtain the form of the general Thomas-Fermi density $n(r)$ given as

\begin{widetext}
\begin{align}
n(r) = -\frac{4}{3  \tilde{C}^{2} r^{2}}
\left\{
 -\bigl[\mathrm{g} - \Omega \tilde{C}\, r^{2}\bigr] 
 + \sqrt{
   \bigl[\mathrm{g} - \Omega \tilde{C}\, r^{2}\bigr]^{2}
   + \frac{3  \tilde{C}^{2} r^{2}}{2}
   \left[
     V(r)
     - \frac{1}{2} 
       \left(r^2\Omega^{2} - \frac{\nu^{2}}{r^{2}}\right)
     - \mu
   \right]
 }
\right\}.
\label{eq:nr}
\end{align}
\end{widetext}

In the presence of the ring potential, the Thomas-Fermi density defined by Eq.~\eqref{eq:nr} possesses the non-trivial turning points (radii) $R_{\rm TF}$ which can be obtained from the solutions of the \me{quartic} equation
\begin{align}\label{eqn:tfrad}
	\frac{\nu^2}{2R_{\rm TF}^2}+\frac{1}{2}\big(R_{\rm TF}-r_0\big)^2-\frac{1}{2}R_{\rm TF}^2\Omega^2-\mu=0.
\end{align}
for a given chemical potential $\mu$, quantum circulation $\nu$, rigid body rotation $\Omega$ and ring radius $r_0$.

\begin{figure}[!ht]
\centering
\includegraphics[width=\columnwidth]{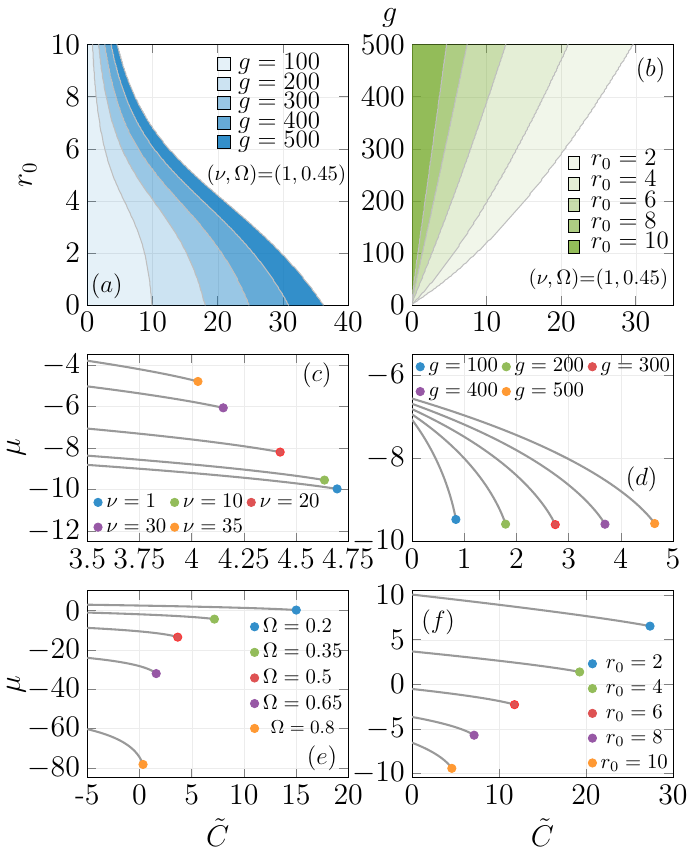}
\me{\caption{Parameter space stability diagram. Panels (a) and (b) show the allowed solution regimes in the $(r_0,C)$ and $(g,C)$ parameter planes respectively. The allowed domains of the chemical potential $\mu$ in terms of the nonlinear rotation strength $C$ are presented in panels (c)-(f). Solution branches are computed for different values of the quantum circulation $\nu$ (c), the van der Waals interaction strength $g$ (d), rigid-body rotation $\Omega$ (e) and ring radius $r_0$.}}
\label{fig:stability}
\end{figure}

\begin{figure}[!b]
\centering
\includegraphics[width=\linewidth]{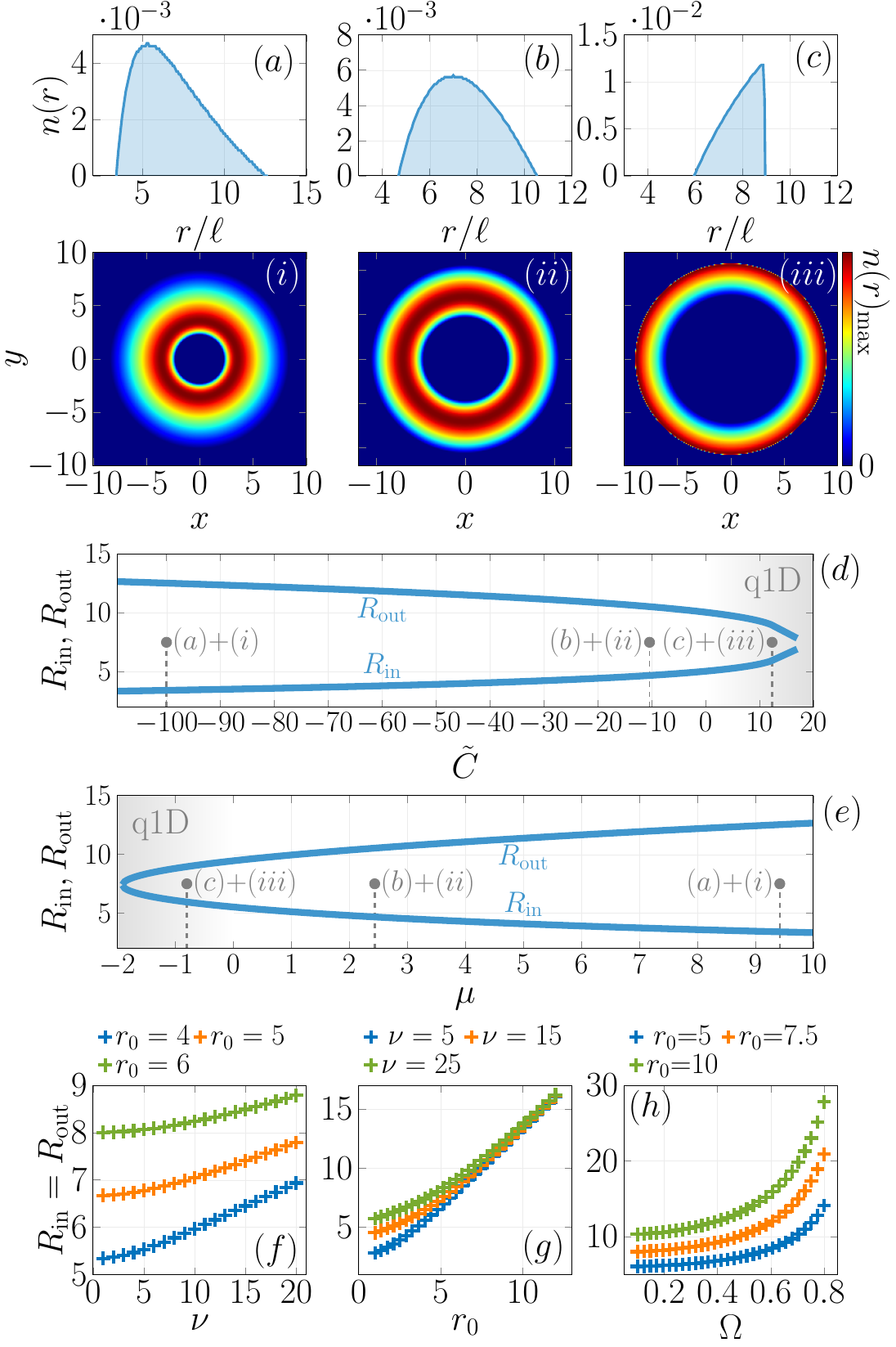}
\me{\caption{Thomas-Fermi profiles. (a)-(c) show example radial density profiles, solutions to Eqs.~\eqref{eqn:tfrad} and \eqref{eqn:tfnorm} with corresponding cartesian densities in panels (i)-(iii). These solutions are indicated within the parameter space for the nonlinear rotation strength $C$ and chemical potential $\mu$ in (d) and (e). Meanwhile panels (f)-(h) explore the behaviour of the point at which the ring disappears $R_{\rm in}=R_{\rm out}$ for different combinations of the model's parameters.}}
\label{fig:tfermi}
\end{figure}

Equation~\ref{eq:nr} demonstrates that the Thomas-Fermi density is influenced significantly by the strength of the nonlinear rotation~$\tilde{C}$ and the radius of the toroidal potential~$r_0$. Specifically, Eq.~\ref{eq:tf_den} illustrates that the centrifugal force increases with $\tilde{C}$, which works against confinement. Therefore, depending on the values of $r_0$ and $\tilde{C}$, the condensate may exceed the confinement. To determine the admissible solution region and find the maximum value of $\tilde{C}$ that maintains a stable condensate, we impose the normalization condition 
\begin{equation}\label{eqn:tfnorm}
2\pi \int\limits_{\me{R_{\text{in}}}}^{\me{R_{\text{out}}}} r\, n(r)\, dr = 1,
\end{equation}
where the inner and outer radii \me{$R_{\text{in}}$} and \me{$R_{\text{out}}$} are the points at which the condensate density vanishes ($n(r) = 0$). 
\me{To understand the full parameter space of the nonlinear model, we can solve Eqs.~\eqref{eqn:tfrad} and \eqref{eqn:tfnorm} in order to obtain the allowed solutions for the toroidally-confined ground states of Eq.~\eqref{eq:dmless_eq} in the Thomas-Fermi limit.} 

\me{Figure \ref{fig:stability} explores the numerically obtained solutions in the Thomas-Fermi limit for different parameter choices. Panel (a) shows the allowed solution boundaries in the $(r_0,\tilde{C})$ plane, with $(\nu,\Omega)=(1,0.45)$. Increasing the van der Waals interaction strength $g$ has the effect of increasing the region in this plane in which the Thomas-Fermi condensate is stable, while toroidal rings with larger radii $r_0$ reduce the allowed solution region. Then panel (b) shows the allowed solutions instead in the $(g,\tilde{C})$ parameter plane. Individual shaded regions correspond to different choices of $r_0$. Increasing $g$ has the general effect of increasing the region of the solution space available to Thomas-Fermi solutions, while smaller rings support solutions in a broader region of $(g,\tilde{C})$, similar to panel (a). Next panels (c)-(f) explore how the chemical potential $\mu$ behaves as the strength of the nonlinear rotation $\tilde{C}$ changes. In panel (c) we fix $g=500$, $\Omega=0.45$ and $r_0=10$. Then, the allowed solutions for $\mu$ are computed for different fixed values of $\nu$. The maximum allowed value of $\tilde{C}$ at which the solutions terminate (coloured circles) decreases as the quantum circulation $\nu$ increases. Following this, panel (d) presents the solutions for $\mu$ with fixed $\nu=10$, $\Omega=0.45$ and $r_0=10$. Here the maximum permissible Thomas-Fermi solution is found to increase as $g$ increases, which might be expected as the size of the condensate scales generally with the van der Waals interaction strength. The next panel (e) explores the effect of changing the strength of the rigid-body rotation $\Omega$, here we fix $\nu=1$, $g=500$ and $r_0=10$. The effect of increasing $\Omega$ moves the maximum allowed solution of $\mu$ to smaller values of $\tilde{C}$, as well as causing the solution region of $\mu$ to broaden. Indeed, for larger $\Omega$ the solutions move to more attractive values of $\mu$. The final panel (f) of Fig.~\ref{fig:stability} presents solutions for $\mu$ for different $r_0$ with fixed $\nu=10$, $g=500$ and $\Omega=0.45$. Here, a qualitatively similar behaviour to panel (e) is observed, where increasing $r_0$ causes the solution to terminate at smaller values of $\tilde{C}$. The effect of the nonlinear rotation moves the solutions from the repulsive to the attractive regime, hence decreasing the domain over which the nonlinear rotation can support stable Thomas-Fermi profiles.}

\me{To understand the nature of the Thomas-Fermi solutions further, Fig.~\ref{fig:tfermi} depicts example solutions in panels (a)-(c) and (i)-(iii). Here we fix the quantum circulation $\nu=15$, the van der Waals interactions $g=500$, the rotation strength $\Omega=0.5$ and the ring radius $r_0=5$. Then the coupled equations given by Eqs.~\eqref{eqn:tfrad} and \eqref{eqn:tfnorm} along with the density profile, Eq.~\eqref{eq:nr} can be iteratively solved to procure the complete parameter space of Thomas-Fermi density distributions for a given pair of chemical potential and nonlinear rotation strengths $(\mu,\tilde{C})$. The inner $R_{\rm in}$ and outer $R_{\rm out}$ radii of the Thomas-Fermi density are shown in panels (d) and (e) respectively. While the solutions extend generally for repulsive $\mu>0$ ($\tilde{C}<0$), the solution terminates in the attractive regime $\mu<0$ ($\tilde{C}>0$). To understand this, several example cartesian densities $n(x,y)$ are shown in panels (i)-(iii), with corresponding radial profiles $n(r)$ shown directly above for each case in panels (a)-(c). Panels (a) and (i) are associated with a negative $\tilde{C}\approx-100$ (repulsive interactions). The effective of the nonlinear rotation is to cause an asymmetric distortion of the Thomas-Fermi profile, with atoms preferentially situated close to $r=r_0$. Increasing the strength of the nonlinear rotation to $\tilde{C}\approx-10$, we observe that the shape of the atoms density profile changes to an approximately symmetric configuration, reminiscent of the Thomas-Fermi profile associated with a standard toroidally confined single-component non-rotating Bose gas. This situation is shown in panels (b) and (ii). Then profiles (c) and (iii) show example density profiles for $\tilde{C}\approx10$ (attractive interactions), approaching the end of the allowed solutions on for $\mu<0$. The ring profile has an almost triangular shape, as seen in (c). Here, the width of the ring profile is significantly reduced, and the system is approaching the quasi one-dimensional limit. The gray shaded regions in panels (d) and (e) indicate this limit. The last row of panels (f)-(h) compute the point at which $R_{\rm in}=R_{\rm out}$ in general. This data is obtained by setting $\partial \mu/\partial R_{\rm TF}=0$ which gives a quartic equation for the Thomas-Fermi radius, which can be solved numerically to obtain real positive-valued radii. By combing this expression with Eq.~\eqref{eqn:tfrad} the cubic term in $R_{\rm TF}$ can be eliminated, resulting in a biquadratic expression for $R_{\rm TF}$ with solution}
\me{\begin{equation}\label{eqn:biquad}
R_{*}^2=\frac{-(2\mu-r_{0}^2)\pm\sqrt{(2\mu-r_{0}^2)^2+12\nu^2(1-\Omega^2)}}{2(1-\Omega^2)},
\end{equation}}
\me{valid when $0\leq\Omega<1$. Equation \eqref{eqn:biquad} provides an alternative way to calculate the point $R_{*}=R_{\rm in}=R_{\rm out}$ for a given set of parameters. Finally, panel (f) shows the behaviour of $R_{*}$ as the quantum circulation increases for different fixed values of $r_0$, while (g) shows the effect of increasing the radius $r_0$ for different $\nu$. The final panel depicts how changing $\Omega$ influences the point $R_{*}$, which is found to increase, being larger in general for increasing $r_0$.}
\begin{figure}[!t]
		\includegraphics[scale=0.45]{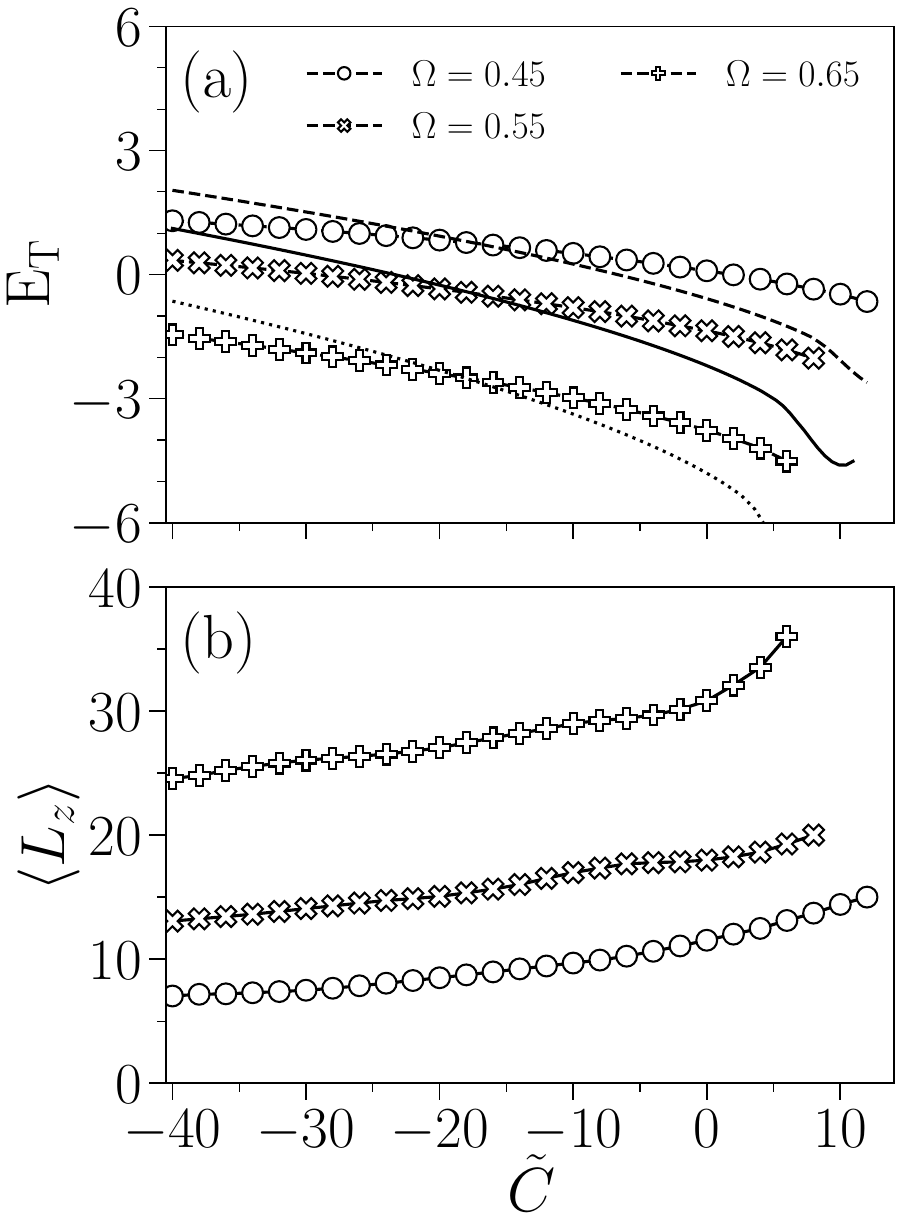}
	\caption{(a) The variation of total energy $\rm E_T$, and (b) angular momentum $\left\langle L_z \right\rangle$ as a function of nonlinear rotation strength $\tilde{C}$, for rotational frequencies  $\Omega=0.45, 0.55$, and $0.65$. The black dashed, dotted, and dash-dotted lines show the energy prediction from the Thomas-Fermi density distribution. The total energy decreases as  $\tilde{C}$ increases, while the angular momentum increases with $\tilde{C}$.}
    \label{fig:energy_ang}
	\end{figure}

\begin{figure}[!t]
		\includegraphics[width=0.75\linewidth]{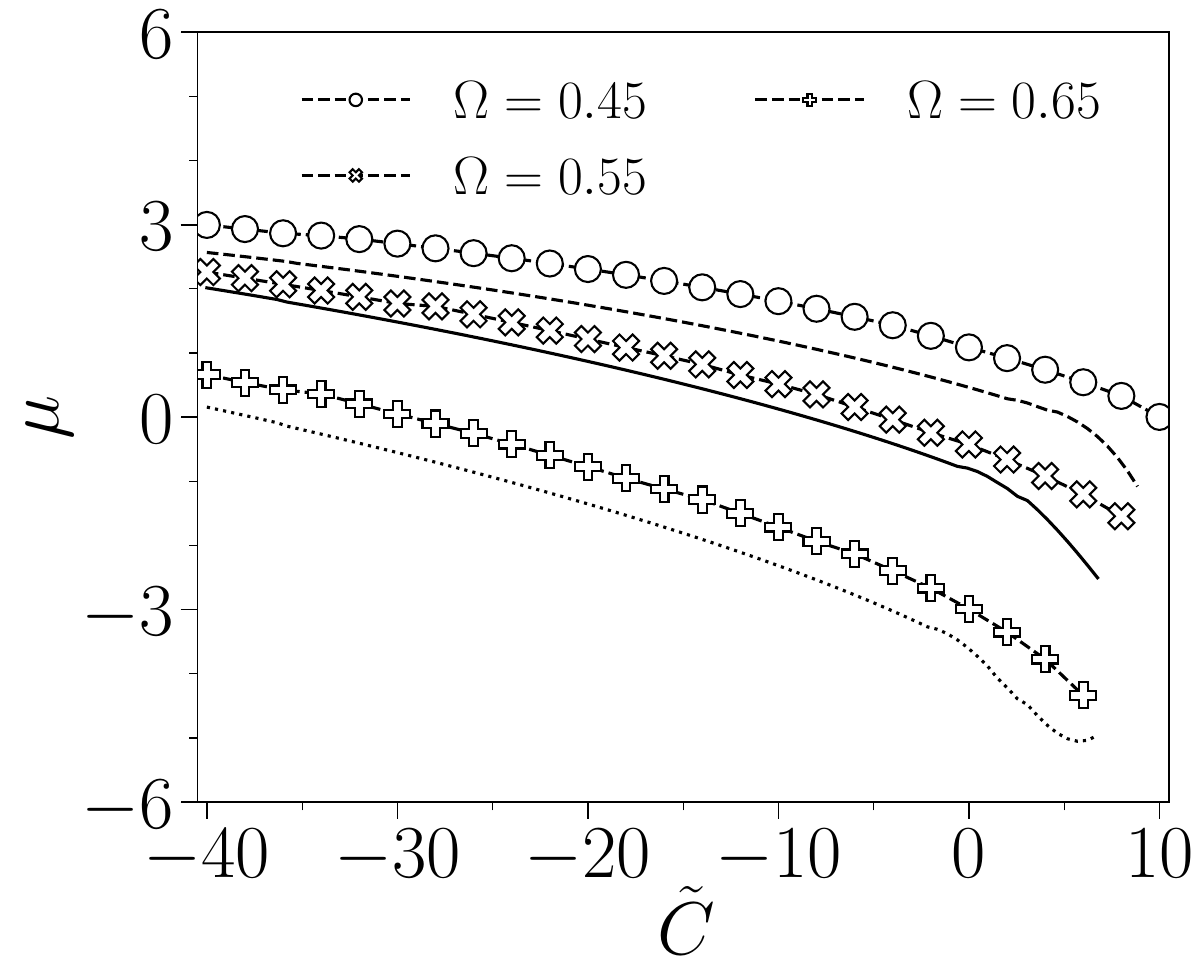}
	\caption{The variation of the chemical potential $\mu$ as a function of the nonlinear rotation strength $\tilde{C}$ for rotational frequencies $\Omega = 0.45, 0.55$, and $0.65$. The black dashed, solid, and dotted lines represent the corresponding predictions from the Thomas--Fermi density distribution.}
    \label{fig:chem}
	\end{figure}

To further characterize the structural transition observed in Fig.~\ref{fig:density}, we calculate conserved quantities in the form of the total energy and the angular momentum as
\begin{subequations} 
\begin{align}
E_T &{=} \int\left(\frac{1}{2}\left\vert\nabla \psi \right\vert^{2}{+}V{+}\frac{\rm g}{2}\left\vert\psi \right\vert^{4} {-} \Omega_n \psi^{*}\hat{L}_z\psi \right) dxdy, \\  
\left\langle L_z\right\rangle &= -i\int\psi^{*}\bigg(x\frac{\partial}{\partial y}-y\frac{\partial}{\partial x}\bigg)\psi dxdy. 
\end{align}
\end{subequations}
We then examine their variation as a function of the nonlinear rotation strength $\tilde{C}$, as shown in Fig.~\ref{fig:energy_ang}. We find that as $\tilde{C}$ increases from negative to positive values, the energy of the system decreases, indicating the energetically more favorable state in case of giant vortex in compare to ring vortex lattice structure. A similar decrease in energy associated with the formation of a giant vortex has been reported in the literature for a Mexican-hat potential, where the structural phase transition depends on the angular velocity~\cite{Fetter:2005}. On the other hand, depending on $\tilde{C}$, the expectation value of the angular momentum, $\langle L_z \rangle$, increases monotonically due to the modulation of the condensate background density and the associated rearrangement of the vortices. 

We find that stable vortices exist in the ring-trapped condensate in the presence of nonlinear rotation. This is shown in Fig.~\ref{fig:chem}, where the chemical potential $\mu <0 $. The presence of attractive interactions in quasi two-dimensional condensates can lead to the collapse of the condensate's wave function when the focussing nonlinearity overwhelms the kinetic energy. On the other hand, the condensate can also be destabilized when the net rotation exceeds that of the confining potential. It has however been shown that vortices may exist in models with attractive interactions, for example in quantum droplets \cite{Andre:2018,Tengstrand:2019,Li:2018}. There are also recent studies of vortices in the presence of systems supporting different cubic nonlinearities where one of the nonlinear terms contributes an attractive component to the overall repulsive system \cite{Sabari:2025,Banerjee:2025}.
In our study, the stability of the vortices is supported by the ring-shaped trapping geometry, which effectively acts as a pinning potential for the giant-vortex state by providing a low density central region. Furthermore, the nonlinear rotation induces an enhanced Magnus force on the vortices confined within the annular region for $\tilde{C}>0$, driving them toward the trap center and thereby promoting the clustering of circulation quanta into a stable giant-vortex configuration. We find that the combined effect of the ring confinement and nonlinear rotation plays a crucial role in stabilizing the giant-vortex states in the attractive regime.

\begin{figure*}
 \centering
 \includegraphics[width=0.9\linewidth]{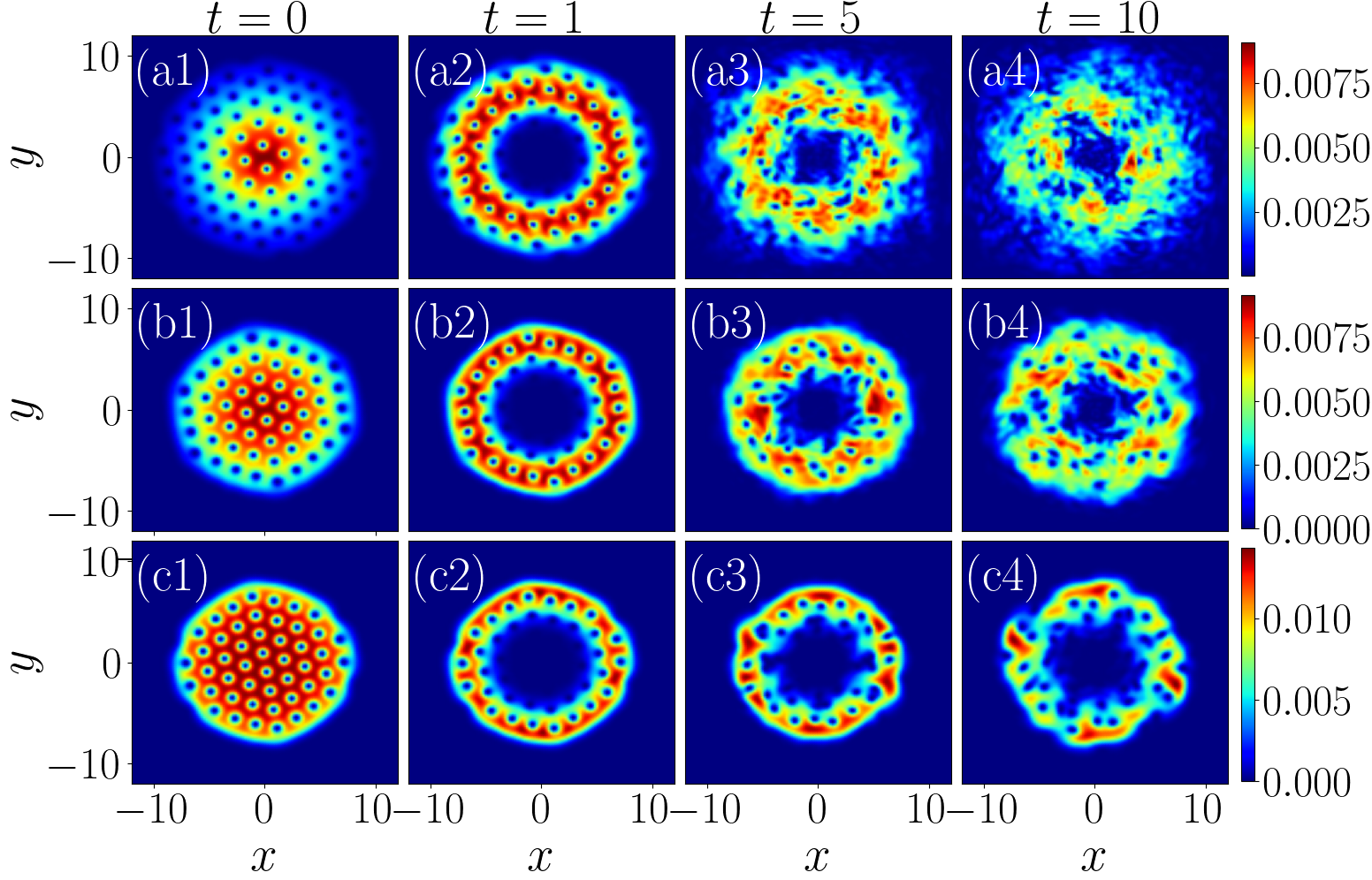}
 \caption{Real-space condensate density profile at different time snaps illustrating the formation of a giant vortex centre, induced by an atom-removal Gaussian perturbation applied to the ground state at different values of $\tilde{C}$ at $\Omega=0.9$ and $g=1000$. (a1)–(a5): for fixed $\tilde{C} = -30$ and $t = (0, 1, 5, 10)$;  (b1)–(b5): for fixed $\tilde{C}= 0$ and $t =  (0, 1, 5, 10)$; (c1)–(c5): for fixed $\tilde{C} = 16$ and $t =  (0, 1, 5, 10)$. }
 \label{fig:density_absorb_pot}
 \end{figure*}

\begin{figure}[!htp]
 \centering
 \includegraphics[width=\linewidth]{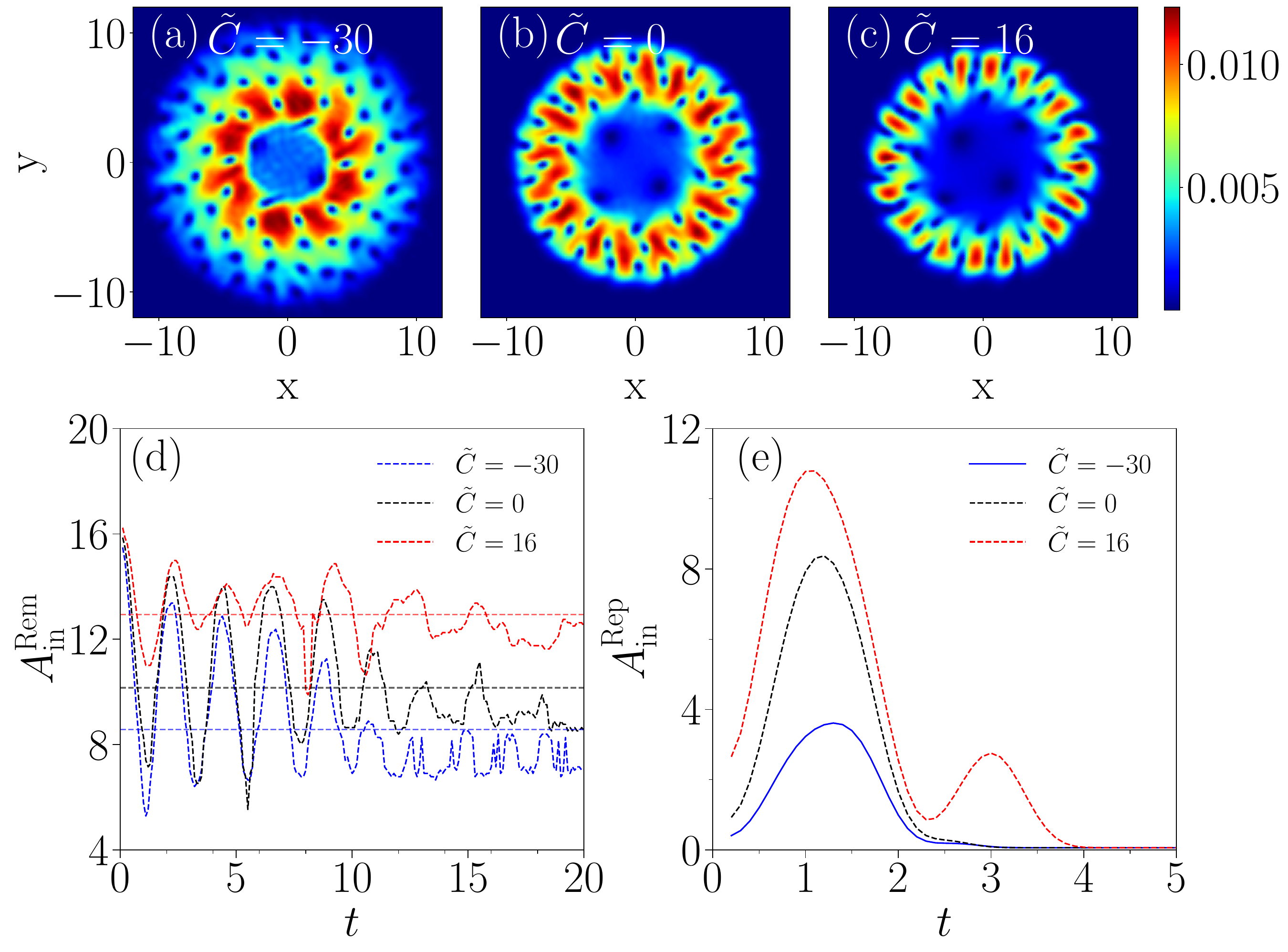}
 \caption{(a)-(c) show the Pseudo colour presentation of the condensate density at $t=1$ for three different values of $\tilde{C}$, illustrating the emergence of a giant vortex at the center due to the repulsive Gaussian perturbation with $V_0 = 100$. (d)
 The time evolution of the core area $A_{\rm in}^{\rm Rem}$ for different $\tilde{C}$, when atoms are removed from the center. The dashed horizontal lines indicate the mean value of the core area oscillations. (e) The time evolution of the core area $A_{\rm in}^{\rm Rep}$ for different values of $\tilde{C}$ in presence of repulsive potential. The core area attains its maximum for $\tilde{C}=16$ and minimum for $\tilde{C}=-30$.}
 \label{fig:core_area_absorb_pot}
 \end{figure}



\section{Giant Vortex Formation: absorptive and repulsive technique}
\label{sec:absorb_repulse}
In addition to toroidal confinement, a giant vortex structure can also be generated in a rapidly rotating condensate trapped in a harmonic potential via two alternative methods: atom removal and a repulsive pinning potential~\cite{Simula:2004,Engles:2003,Simula:2005}. For the case of atom-removal, a resonant laser beam is focused on the condensate center to create a density depletion. The resulting mean field pressure gradient attempts to drive atoms toward the hole, but the centrifugal potential associated with rotation deflects the inflowing atoms and instead generates a circulating flow around the depletion, which gives rise to a giant vortex and the toroidal geometry. On the other hand, a repulsive (pinning) potential at the condensate center pushes individual vortex cores outward and forces them to arrange around the barrier and thereby create a ring-shaped density. In the following section, we incorporate the laser-induced potential into the GPE~(Eq.~\ref{eq:dmless_eq}) and examine the impact of gauge-induced nonlinear rotation on the formation of a giant vortex. To consider the effect of the laser beam, we introduce a Gaussian perturbation in the trap, so the  effective potential becomes 
\begin{subequations}
\begin{align}
V(x,y)&=\frac{1}{2}(x^2+y^2)+V_{\text{laser}}, \\
V_{\rm {laser}}&=V_0\exp\left[-\frac{(x-x_0)^2+(y-y_0)^2}{(r_0/2)^2}\right]
\end{align}
\end{subequations}
where $V_{\rm laser}$ is the Gaussian perturbation with barrier height $V_0$, beam center ($x_0, y_0$) and width $r_0$. The beam power $P=\int dr \ V_{\rm laser}^{2}$ is directly related to the induced hole and the subsequent modification of the condensate density distribution. Higher values of $P$ correspond to more intense Gaussian beam perturbations, which result in greater density depletion near the trap center and the formation of a larger low density hole. The generalized GP equation is then evolved in real-time, using the stationary state obtained at $\Omega=0.9$ and $\rm g=1000$ as the initial state.
In the present work, we consider an imaginary $V_0=-100i$  to implement the  atom removal potential  with the other parameters $r_0=9$, and $(x_0 =0, \ y_0=0)$. In the simulation the potential is applied over a dimensionless time interval of $t=[0, 0.05]$. \\


To systematically examine the effect of the nonlinear rotation on giant-vortex formation using the atom removal technique, in Fig.~\ref{fig:density_absorb_pot}, we present snapshots of condensate density at $t=0,  1, 5$, and $10$, where the first~[(a1)-(a5)], second~[(b1)-(b5)], and third~[(c1)-(c5)] rows correspond to $\tilde{C}=-30$, $0$, and $16$, respectively. We find that at $t=0$ the condensate forms a hexagonal vortex-lattice structure for $\tilde{C}=0$, and for $\tilde{C}=-30$ the vortices are pushed to the periphery of the condensate, whereas for $\tilde{C}=16$ they concentrate near the center of the trap. This behavior is associated with $\tilde{C}$ and can be attributed to the Magnus force \me{which} has been studied previously~\cite{Edmonds:2020,Bhat:2023,Bhat:2024}. 
Interestingly, at $t = 1$, the central density becomes strongly suppressed, leading to the formation of a giant vortex irrespective of the nonlinear rotation strength $\tilde{C}$. We further observe that the inner radius of the toroidal density increases for positive $\tilde{C}$ and decreases for negative $\tilde{C}$; correspondingly, the annular condensate becomes narrower for larger $\tilde{C}$. Time snapshots also show that the hole area oscillates with time.
To characterize these radial motions more clearly, in Fig.~\ref{fig:core_area_absorb_pot}(d) we show the variation of the core area with time for $\tilde{C} = -30$, $0$, and $16$. The mean core area is largest for $\tilde{C} = 16$ and smallest for $\tilde{C} = -30$. This behaviour arises from the interplay between the mean-field pressure gradient, which tends to refill the density hole, and the centrifugal potential, which acts outward. The variation in the mean core area can be understood from the centrifugal potential in Eq.~\ref{eq:tf_den}, whose strength increases for positive $\tilde{C}$ and decreases for negative $\tilde{C}$.

Following the analysis of atom-removal dynamics, we now extend our study to investigate the effect of nonlinear rotation on the formation of a giant vortex in the presence of a repulsive Gaussian potential. 
Here we consider a repulsive Gaussian potential with a barrier height of $V_0 = 100$ and apply it for the same interval as the atom removal. All other parameters remain consistent with the previous analysis. 
In Fig.~\ref{fig:core_area_absorb_pot}(a)-(c), we present the condensate density at $t=1$ for values of $\tilde{C} = -30$, $0$, and $16$. The density profile shows that, similar to the previous case, a giant vortex forms at the center, with the radius of this vortex being larger for $\tilde{C} > 0$ compared to when $\tilde{C} < 0$. The stripes observed in the density depend on the amplitude of the laser beam and have been studied previously in the literature~\cite{Simula:2005}. 
Figure~\ref{fig:core_area_absorb_pot}(e) shows the time evolution of the giant-vortex core area. Consistent with the density profiles, the core area is larger for $\tilde{C}>0$ than for $\tilde{C}<0$. After some time, the core area decreases to zero as atoms from the outer region refill the depleted center, causing the giant vortex to collapse.
However, no oscillation has been observed in the core area for the repulsive potential. The reason for this is that the angular momentum of a rotating vortex ring with radius $R$ is proportional to $L \propto \Omega_{n} R^2$. When  atoms are removed from the center, the angular momentum of the atoms within the annular region increases, resulting in a new lattice with fewer atoms and a higher angular velocity $\Omega$.  As atoms in the outer region flow inward to refill the depletion under the action of the mean-field pressure, they encounter an effective barrier created by this increased angular momentum: the condensate cannot accelerate its rotation as fast as the hole fills and leads to an oscillation of the core area. 
In contrast, when a repulsive potential is applied, the total angular momentum of the condensate is conserved and no net increase of angular momentum per particle is induced in the annulus. Consequently, once the perturbation is removed, the system relaxes and atoms from the outer region refill the central hole. Since the angular-momentum distribution is not significantly altered, no oscillations of the core area are observed in this case.

\section{Collective excitation spectrum}
\label{sec:collective}
After establishing the dependence of the ground state configuration on the nonlinear rotation, particularly the variation in the size of the central hole, we now examine the collective excitation spectrum. Two complementary approaches are used to investigate the low-lying modes. The first approach utilizes the Bogoliubov-de Gennes (BdG) framework, which constitutes a linear stability analysis of the stationary condensate solution and yields the full quasiparticle spectrum. The second approach is the hydrodynamic method, derived from the Madelung transformation of the Gross-Pitaevskii equation, which describes the system in terms of density and phase fluctuations. Both methods are applied to determine the excitation spectrum as the strength of the nonlinear rotation is varied.

Within the Bogoliubov-de Gennes (BdG) framework, the stationary ground state $\psi_{0}$ is perturbed by a small fluctuation $\delta\psi$, resulting in the condensate wave function being expressed as
\begin{align}
\psi(x,y,t)=e^{-i\mu t}\!\left[\psi_{0}(x,y)+\delta\psi(x,y,t)\right],
\label{eq:pertrb_psi}
\end{align}
where $\mu$ is the chemical potential of the ground state. The fluctuation $\delta\psi$ is expanded in terms of quasi particle amplitudes as
\begin{align}
\delta\psi(x,y,t)=u(x,y)e^{i\omega t}-v^{*}(x,y)e^{-i\omega t},
\label{eq:delpsi}
\end{align}
In this context, $u(x,y)$ and $v(x,y)$ represent the Bogoliubov mode functions, while $\omega$ denotes the corresponding excitation frequency. These amplitudes satisfy the following normalization condition:
\begin{align}
\int\int \left(|u(x,y)|^{2}-|v(x,y)|^{2}\right)\,dx\,dy = 1.
\label{eq:bdg_norm}
\end{align}

By substituting Eqs.~\eqref{eq:pertrb_psi} and \eqref{eq:delpsi} into the time dependent Gross–Pitaevskii (GP) equation, (Eq.~\eqref{eq:dmless_eq}), and retaining only terms linear in $\delta\psi$, the coefficients of $e^{i\omega t}$ and $e^{-i\omega t}$ are equated. This procedure yields the coupled Bogoliubov–de Gennes equations
\begin{subequations}\label{eq:bdg}
	\begin{align}
		\omega u=&\left[\mathcal{L}-\Omega L_z-\tilde{C}(\rvert\psi_0\lvert^{2}L_z+\psi_{0}^{*}L_z\psi_{0})\right]u \notag \\ &
 -[\text{g}\psi_{0}^{2}-\tilde{C}\psi_{0}L_z\psi_0]v,\\ 
		-\omega v=&\left[\mathcal{L}+\Omega L_z+\tilde{C}(\vert \psi_{0}\vert ^{2}L_z+\psi_{0}L_z\psi_{0}^{*})\right]v \notag \\ &
 -[\text{g}\psi_{0}^{*2}+\tilde{C}\psi_0^{*}L_z\psi_0^{*}]u,
	\end{align}
\end{subequations}
in Eqs.~\eqref{eq:bdg} we define the function
\begin{align}
\mathcal{L}= -\frac{1}{2}\nabla^2+V+2\text{g}\rvert\psi_0\lvert^{2}-\mu.
\end{align}
The BdG equations are obtained by linearizing the Gross-Pitaevskii equation around the stationary solution using the perturbation ansatz given in Eq.~(18). In this formulation, the kinetic-energy operator remains $-\nabla^2/2$, while the effect of the quantum circulation $\nu$ is already incorporated into the stationary condensate wavefunction $\psi_0$, and consequently into the equilibrium density $|\psi_0|^2$. Therefore, the centrifugal contribution, $\nu^2/(2r^2)$, is implicitly included through the background solution and does not appear as an additional explicit potential term in the operator.
Solving Eqs.~\eqref{eq:bdg} provides the excitation frequencies $\omega$ together with the corresponding quasi particle modes $u$ and $v$.




Another approach to finding the frequencies of collective modes is through the hydrodynamic method. In this approach, the time-dependent GP equation~(Eq.~\ref{eq:dmless_eq}) is recast in terms of the condensate density and phase via the Madelung transformation. This technique has been widely used to analyze the collective excitations of annular condensates~\cite{Cozzini:2005,Cozzini:2006}. Using a similar method, we derive the equations for the density $n$ and phase $S$ that are suitable for our system. We then linearize these equations around the equilibrium state by introducing small fluctuations in phase $\delta S$, and density $\delta n $, which leads to the following set of linearized hydrodynamic equations.


\begin{subequations}\label{eq:hydro}
\begin{align}
\frac{\partial \delta n}{\partial t}
+\left[\frac{\nu}{r^{2}}-(\Omega+\tilde{C}n_{0})\right]
\frac{\partial \delta n}{\partial \phi}
+\nabla\cdot\!\left(n_{0}\nabla\delta S\right)
&=0,\\
\frac{\partial \delta S}{\partial t}
+\left[\frac{\nu}{r^{2}}-(\Omega+\tilde{C}n_{0})\right]
\frac{\partial \delta S}{\partial \phi}
+\left(g-\tilde{C}\nu\right)\delta n
&=0.
\end{align}
\end{subequations}

\begin{figure}[!t]
 \centering
 \includegraphics[scale=0.35]{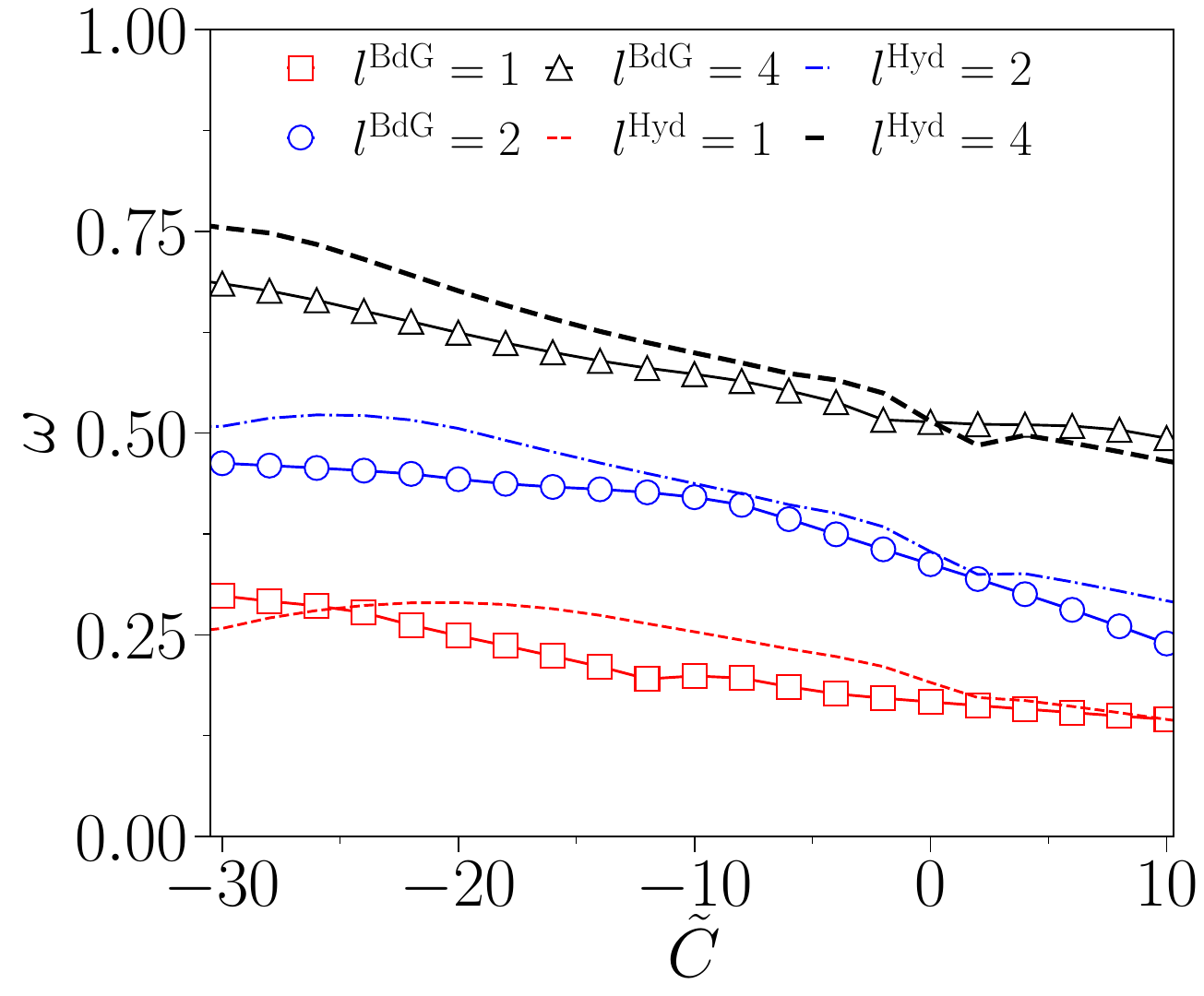}
 \caption{The collective excitation spectrum for different angular quantum numbers $l$ as a function of nonlinear rotation strength $\tilde{C}$ at $ \Omega = 0.45$ and ${\rm g} = 250$. The symbols represent the frequencies obtained from the BdG equation, while the dashed lines illustrate the frequencies of the normal modes obtained from hydrodynamic theory.}
 \label{fig:excitation_spec}
 \end{figure}
where $n_0$ is the equilibrium Thomas-Fermi density, and can be obtained from ~Eq.~\eqref{eq:nr}. Then by considering the density and phase modulations of the form $\delta n$ and $\delta S \propto \exp(i(l\phi - \omega t))$, we derive the following set of equations.

\begin{subequations}\label{eq:hydro_lin}
\begin{align}
\left(g-\tilde{C}\nu\right)\delta n -i\left[\omega-\frac{l\nu}{r^2}+l(\Omega+\tilde{C}n_0)\right]\delta S
=0,\\
\left[-\omega+\frac{l\nu}{r^2}-l(\Omega+\tilde{C}n_0)
\right]i\delta n+\frac{1}{r}\frac{\partial}{\partial r}
\left(rn_0\frac{\partial\delta S}{\partial r}\right)\\
-\frac{l^2n_0}{r^2}\delta S &=0.
\end{align}
\end{subequations}
After substituting Eq.~\ref{eq:hydro_lin}(a) into the Eq.~\ref{eq:hydro_lin}(b), we obtain the following equation:
\begin{align}\label{eq:hydro_eig}
&
\left[\left(\omega-\frac{l\nu}{r^2}+l(\Omega+\tilde{C}n_0)
\right)^2-\left(g-\tilde{C}\nu\right)\frac{l^2n_0}{r^2}\right]\delta S
\notag\\
&
+\left(g-\tilde{C}\nu\right)\frac{1}{r}\frac{\partial}{\partial r}
\left(rn_0\frac{\partial\delta S}{\partial r}\right)=0.
\end{align}
We solve Eq.~\eqref{eq:hydro_eig} numerically to obtain the collective mode frequencies, imposing zero-flux boundary conditions and discretizing the annular domain $[R_{\rm in}, R_{\rm out}]$ using $N$ interior grid points.

In Fig.~\ref{fig:excitation_spec}, we show the dependence of the low-lying excitation frequencies on the nonlinear rotation strength $\tilde{C}$ for fixed rotation frequency $\Omega=0.45$ and interaction strength $\rm g=250$.
\begin{figure}[!t]
 \centering
 \includegraphics[scale=0.35]{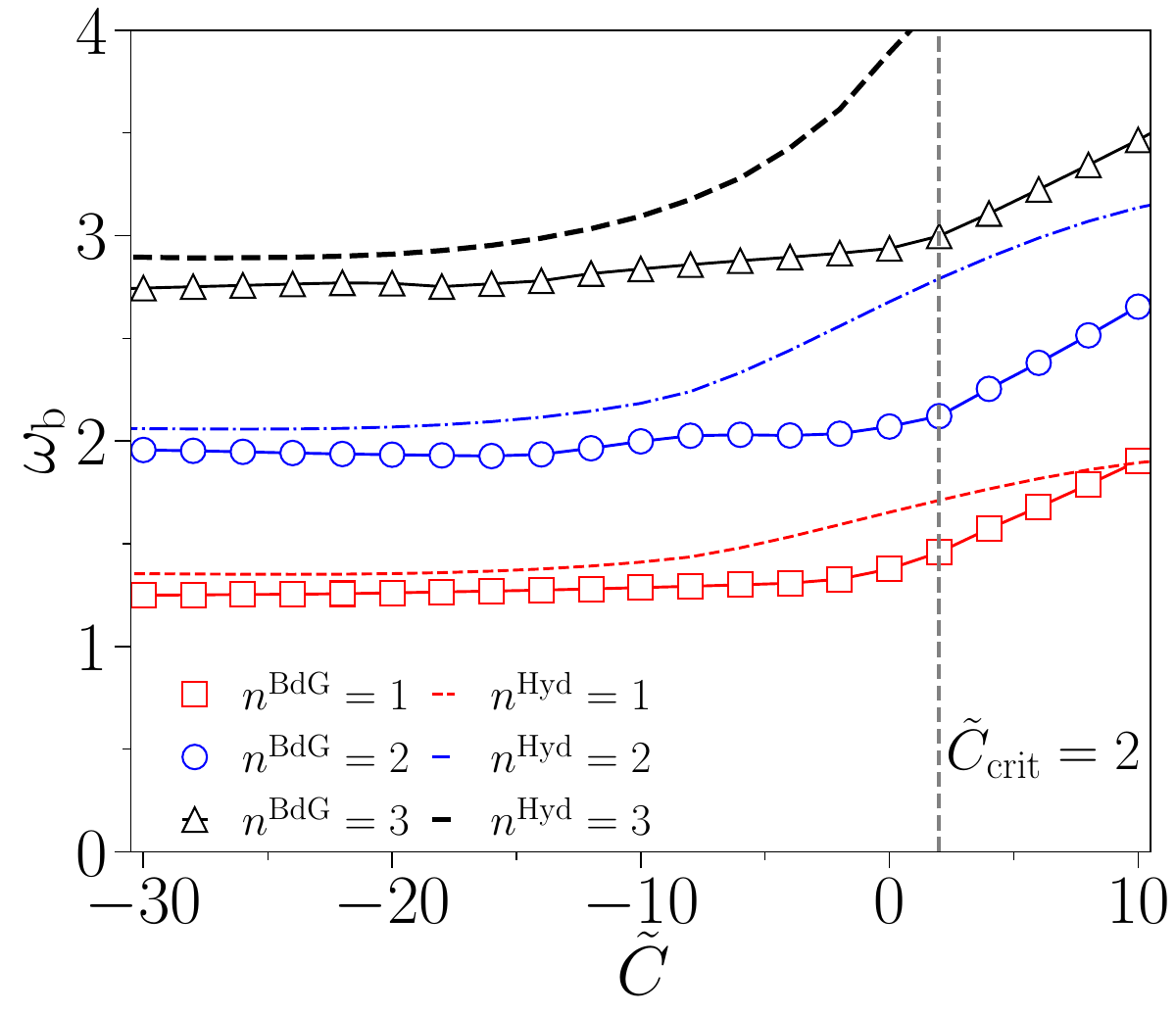}
 \caption{Variation of the breathing mode frequency ($l=0$) with nonlinear rotation strength $\tilde{C}$ for three lowest radial quantum numbers $n=1, 2,$ and $3$. The markers and dashed lines represent the breathing mode frequencies obtained from the BdG and hydrodynamic approaches, respectively. The deviation of $\rm \omega_b$ beyond a critical value of $\tilde{C}$ indicates a transition from a ring shaped vortex state to a giant vortex state.}
 \label{fig:breath_freq}
 \end{figure}

The frequencies obtained from the hydrodynamic approach by solving Eq.~\ref{eq:hydro_eig} are shown as dashed curves, while the symbols denote the corresponding BdG results.  
The excitation branches corresponding to the azimuthal quantum numbers $l=1$, $2$, and $4$ all decrease with increasing $\tilde{C}$, indicating a softening of the spectrum under stronger nonlinear rotation. In particular, the decrease in the dipole mode ($l=1$) signals a deviation from Kohn's theorem, which states that the dipole frequency should remain independent of the interaction strength. Similar violations have been reported in harmonically trapped condensate and in spin-orbit-coupled condensates~\cite{Zhang:2012, Chen:2012}. This reduction arises from the coupling between density and angular momentum. The softening of the higher-order modes further indicates a structural crossover of the condensate from a ring-shaped vortex lattice to a giant-vortex state as the nonlinear rotation becomes stronger.  \\

Since the annular width of the condensate depends strongly on $\tilde{C}$, it is natural to examine the breathing modes, which are directly associated with radial expansion and compression. In Fig.~\ref{fig:breath_freq} we present the three lowest breathing modes ($l=0$), corresponding to radial quantum numbers $n=1$, $2$, and $3$, as functions of $\tilde{C}$. Both the BdG and hydrodynamic approaches yield consistent results. In contrast to the azimuthal modes, the breathing frequencies~$\omega_{\rm b}$ remain nearly constant over a wide range of negative $\tilde{C}$, but increase monotonically once $\tilde{C}$ becomes positive. This behavior reflects the same structural transition toward the giant vortex regime.


The hydrodynamic approach is appropriate in the Thomas-Fermi regime, where the interaction energy dominates over the kinetic energy and the condensate density varies smoothly over length scales larger than the healing length. In this limit, the quantum-pressure term can be neglected, and the Gross-Pitaevskii equation reduces to \me{the} superfluid hydrodynamic equations for the density and phase. These equations accurately describe low-lying, long-wavelength collective excitations for which quantum-pressure effects are negligible. Since our work focuses on such low-energy modes in a strongly interacting annular condensate, the hydrodynamic treatment is well justified.

However, the hydrodynamic description becomes less accurate when the condensate density varies rapidly, especially near boundaries, vortex cores, or for higher-order excitations where quantum-pressure effects become significant. In the present study, the characteristic annular width satisfies $w\gg\xi$, where $\xi$ is the healing length, so the Thomas-Fermi approximation remains valid over most of the condensate except in the vicinity of vortex cores. For positive values of the nonlinear rotation strength, $\tilde{C}>0$, the vortices migrate into the central giant vortex, leaving the annular condensate nearly free of vortex cores. Consequently, the equilibrium density in the annulus becomes smoother, and the assumptions underlying the hydrodynamic theory are better satisfied. In contrast, for $\tilde{C}<0$, vortices remain distributed throughout the annulus, producing local density variations that are fully accounted for in the BdG theory but are not explicitly captured within the hydrodynamic description, leading to larger deviations between the two approaches for the angular modes. Similar mismatches between excitation frequencies obtained from the BdG and hydrodynamic approaches have also been reported in Ref.~\cite{Dalfavo:1999}.\\
The behavior of the breathing modes is qualitatively different since they are governed primarily by radial compression and expansion of the condensate. As the radial quantum number increases, the density develops stronger radial variations, making the neglected quantum-pressure contribution more important. Consequently, the discrepancy between the BdG and hydrodynamic predictions increases for higher-order breathing modes, particularly for the $n=3$ mode.

\begin{figure}[!t]
 \centering
 \includegraphics[width=0.95\linewidth]{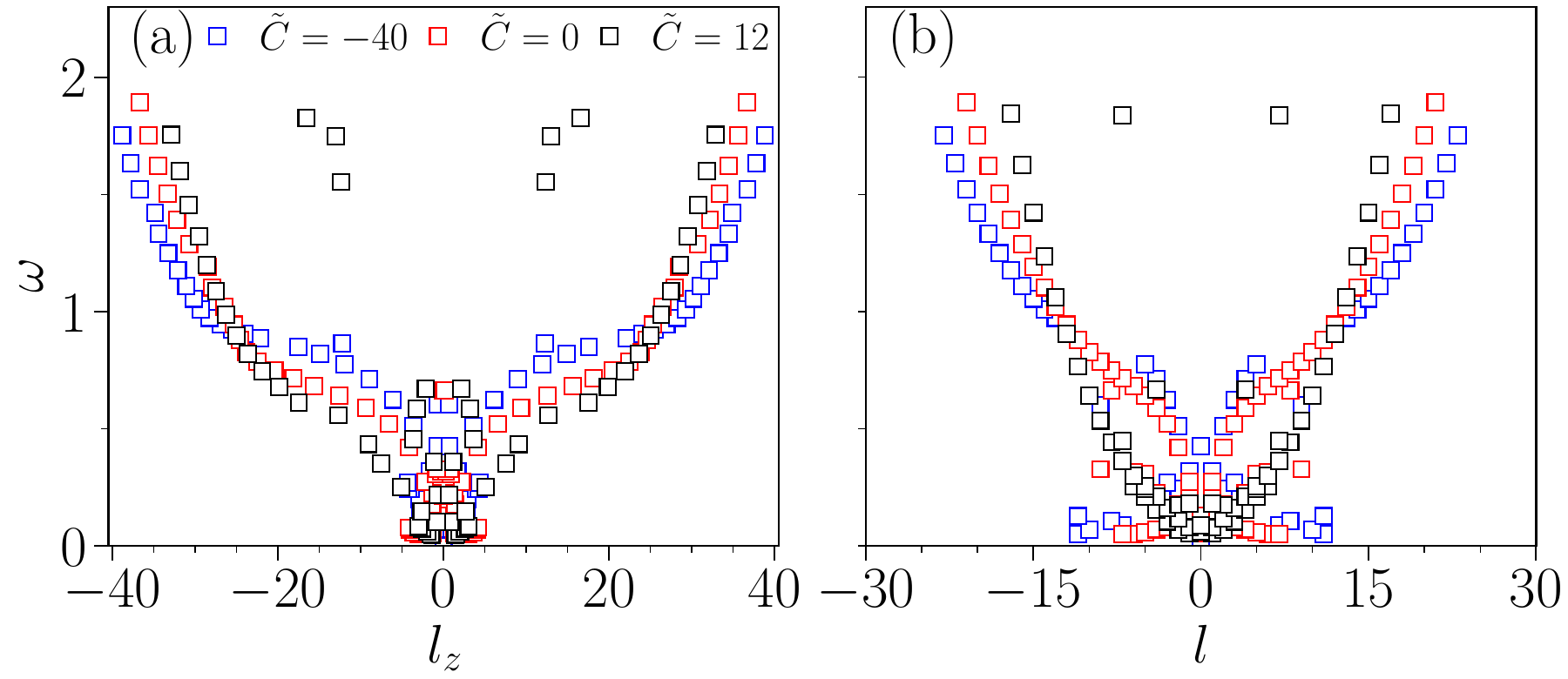}
 \caption{Excitation spectrum as a function of (a) angular momentum, and (b) angular quantum number for $\tilde{C} = -40,0$, and $12$.}
 \label{fig:ang_vs_omg_toroid}
 \end{figure}
\begin{figure*}[!htp]
 \centering
 \includegraphics[width=\linewidth]{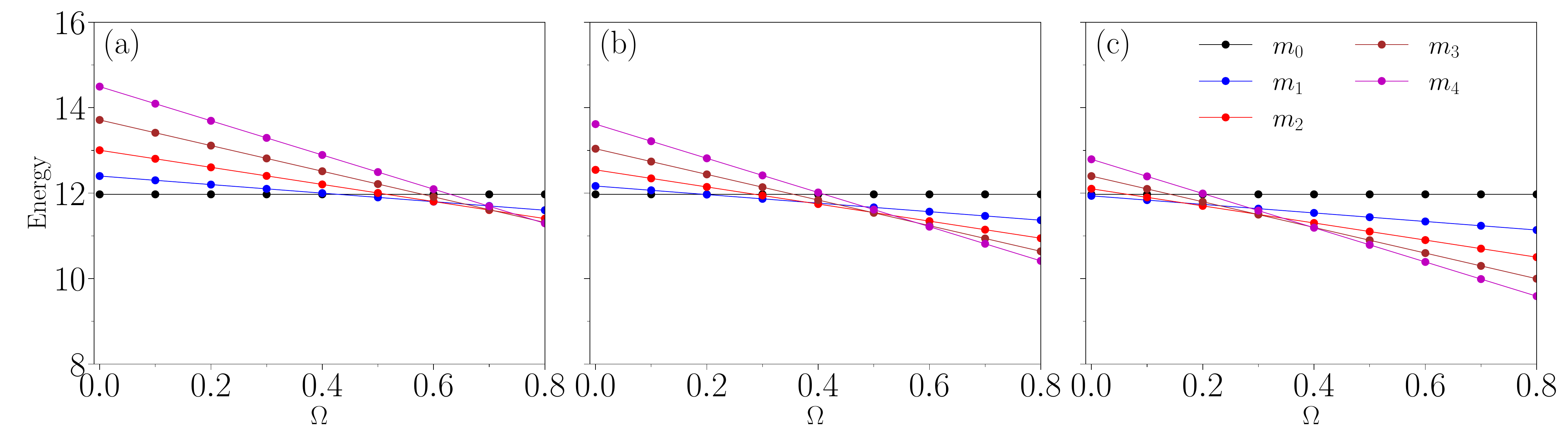}
 \caption{The free energy as a function of the rotation frequency $\Omega$ is shown for nonlinear rotation strengths (a) $\tilde{C}=-40$, (b) $\tilde{C}=0$, and (c) $\tilde{C}=40$. In the absence of nonlinear rotation ($\tilde{C}=0$), the intersection between the free energies of the vortex-free state and the singly quantized vortex occurs at the critical rotation frequency $\Omega_c = 0.21$. For negative nonlinear rotation strength ($\tilde{C}=-40$), this intersection is shifted to higher values of $\Omega$, whereas for positive nonlinear rotation strength ($\tilde{C}=40$), it is shifted toward lower rotation frequencies.
}
 \label{fig:energy_mqv}
 \end{figure*}

\begin{figure*}[!htp]
 \centering
 \includegraphics[width=\linewidth]{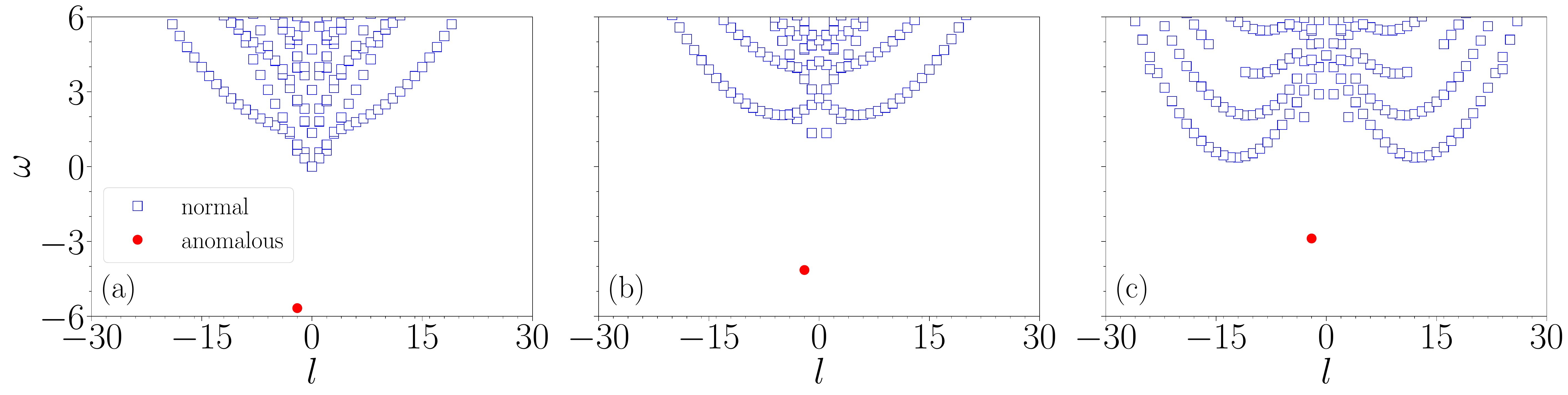}
 \caption{Quasiparticle excitation energies~($\omega$) as a function of the angular quantum number~($l_z$) for a doubly quantized vortex ($m=2$) at a rotation frequency $\Omega=0.2$, illustrated for (a) $\tilde{C}=-40$, (b) $\tilde{C}=0$, and (c) $\tilde{C}=60$. The red marker denotes the anomalous mode corresponding to negative-energy excitations at angular quantum numbers $l=-2$. For negative nonlinear rotation strength ($\tilde{C}=-40$), the anomalous modes shift toward lower energies, indicating an enhanced tendency toward vortex instability and splitting. In contrast, for positive nonlinear rotation strength ($\tilde{C}=60$), the anomalous-mode energies move upward toward zero, suggesting a partial stabilization of the doubly quantized vortex state.}
 \label{fig:excit_mqv}
 \end{figure*}

 
In Fig.~\ref{fig:ang_vs_omg_toroid}(a) and (b), we illustrate the variation of the excitation frequency $\omega$ as a function of angular momentum $l_z$ and the angular quantum number $l$, respectively for three different values of $\tilde{C}$: $\tilde{C} = -40, 0$, and $12$.  The angular momentum of an excitation mode with respect to the rotating condensate is defined by
\begin{align}
l_z=\frac{(l_u - l_\psi)N_u + (l_v + l_\psi)N_v}{N_u + N_v},
\end{align}
where 
\begin{align}
l_\alpha=-i\hbar\frac{\int dxdy\alpha^*\big(x \frac{\partial\alpha}{\partial y} - y\frac{\partial\alpha}{\partial x}\big)}{\int dxdy\alpha^*\alpha},
\end{align}
with $\alpha$ being one of the wave functions $u(x,y)$, $v(x,y)$, or $\psi(x,y)$, and 
\begin{align}
N_u = \int dxdy |u(x,y)|^2, \quad N_v = \int dxdy |v(x,y)|^2.
\end{align}
The spectrum remains symmetric about $l=0$, while the splitting between the $-l$ and $l$ branches depends on the nonlinear rotation parameter $\tilde{C}$. 

\section{Stability of multiply quantized vortices}
\label{sec:stab_mqv}
In this section, we investigate the effect of nonlinear rotation on the stability of multiply quantized vortices (MQVs) in a harmonically trapped condensate. Numerically, MQV states are generated for $\text{g}=400$ by imprinting a higher-charge vortex at the condensate center and subsequently employing the resulting state as the initial guess for imaginary-time propagation. In general, MQVs have higher energy than configurations consisting of singly quantized vortices and therefore tend to dissociate into individual vortices as the system evolves toward lower-energy states. Previous studies have shown that MQVs can be stabilized using a pinning potential, with the stability controlled by the radius and strength of the pinning potential~\cite{Simula:2002}. Such states have also been realized experimentally~\cite{Wilson:2022}.

To examine the influence of nonlinear rotation, we first analyze the free-energy landscape. Figure~\ref{fig:energy_mqv} shows the free energies of MQVs as a function of the rotational frequency $\Omega$ for $\tilde{C}=-40$, $0$, and $40$. For $\tilde{C}=0$ [Fig.~\ref{fig:energy_mqv}(b)], the vortex-free state ($m=0$) has the lowest free energy. As $\Omega$ increases, a critical frequency $\Omega_c$ is reached at which the energy of a singly quantized vortex becomes equal to that of the vortex-free state. Beyond this point, the free energy of higher-charge vortices decreases relative to the vortex-free state.

However, this reduction in free energy does not imply the stabilization of MQVs. For $\Omega>\Omega_c$, angular momentum is typically accommodated through the nucleation of singly quantized vortices from the condensate boundary, leading to nonaxisymmetric vortex configurations and eventually vortex-lattice formation. Consequently, MQVs remain energetically unstable against dissociation into singly quantized vortices.

A qualitatively different behavior is observed for positive nonlinear rotation ($\tilde{C}=40$). In this case, the free-energy crossings between MQV states and the vortex-free state occur at rotation frequencies below $\Omega_c$, where vortex nucleation is energetically unfavorable. As a result, MQVs become the lowest-energy configuration and are therefore globally stable. This stabilization originates from the coupling between the condensate density and angular momentum, which lowers the energy of higher-charge vortex states. In contrast, for $\tilde{C}<0$, this coupling opposes the imposed rotation, shifting the stabilization region to $\Omega>\Omega_c$ and suppressing MQV stability.

To investigate local stability, we next analyze the Bogoliubov--de Gennes (BdG) excitation spectrum. Figure~\ref{fig:excit_mqv} shows the spectrum of a doubly quantized vortex ($m=2$) at $\Omega=0.2$ for $\tilde{C}=-40$, $0$, and $60$. Within the BdG framework, local stability is determined by the presence of anomalous modes, i.e., excitations with negative energy and positive norm ($\omega<0$ and $N>0$), which indicate directions along which the condensate can lower its energy.

For all values of $\tilde{C}$, an anomalous mode appears at angular momentum $l=-2$. This mode is associated with the splitting of the doubly quantized vortex into two singly quantized vortices. Its behavior changes significantly with nonlinear rotation. For $\tilde{C}=-40$, the mode shifts to more negative energies, indicating a strong enhancement of the splitting instability. This occurs because the effective rotational frequency is reduced in the high-density region, weakening the rotational stabilization of the vortex.

For positive nonlinear rotation ($\tilde{C}=60$), the anomalous mode moves upward toward zero energy, indicating a partial suppression of the instability. Nevertheless, the mode remains negative even for this relatively large value of $\tilde{C}$, demonstrating that the doubly quantized vortex is still locally unstable. At larger rotation frequencies and nonlinear rotation strengths, additional negative-energy modes emerge at higher angular momenta, revealing further instability channels.\\
The contrasting behavior for positive and negative nonlinear rotation can be understood from the corresponding modifications of the condensate density and effective rotational strength. For $\tilde{C}=-40$, the central density increases and the vortex core narrows, but the reduced effective rotation enhances the tendency toward vortex splitting. For $\tilde{C}=60$, the density profile becomes more uniform, reducing the density gradients that normally resist vortex splitting. Consequently, although positive nonlinear rotation can lower the free energy of MQVs and make them globally stable, it is not sufficient by itself to eliminate anomalous modes and achieve complete local stability.

\section{Conclusion}
\label{sec:conclusion}
In this work, we have studied the impact of density-dependent gauge potential induced nonlinear rotation on the structural transformation of a vortex lattice in a condensate confined within a toroidal geometry. We find that, depending on the strength of the nonlinear rotation, the system undergoes a transition from a ring shaped vortex lattice to a giant vortex state. In particular, the annular region of the condensate becomes narrower for positive values of the nonlinear rotation strength and wider for negative values. We observe a qualitatively similar influence of nonlinear rotation when the giant vortex state is generated through atom-removal and absorption techniques. In addition, we identify the permissible solution regime by mapping the parameter space defined by the trap radius and the nonlinear rotation strength, which represents the region where the condensate exists. We further examine the effect of nonlinear rotation on the collective excitation spectrum by employing both the BdG and hydrodynamic approaches. Our results reveal a decreasing phenomenon in the dipole mode, indicating a violation of Kohn's theorem. In contrast, the breathing mode exhibits an increasing trend beyond a critical nonlinear rotation strength, signaling radial compression of the condensate. Furthermore, we analyze the influence of nonlinear rotation on the stability of multiply quantized vortices. It was found that while it makes higher order vortices globally stable, it is unable to achieve local stability.

Beyond the analysis of the ground state and excitation spectrum under nonlinear rotation in a toroidal geometry, the problem of persistent current formation presents an interesting direction for future investigation. In particular, it would be worthwhile to explore the impact of the gauge potential induced nonlinear rotation on phase slip dynamics~\cite{MU:2015} and Hysteresis loops~\cite{Jia:2025}. Furthermore, incorporating impurity effects and dissipation could provide deeper insight into the dissipative dynamics and vortex excitations in such systems~\cite{Yakimenko:2015}.

\section{Acknowledgments}
R.B. acknowledges the financial support received from the Department of Science and Technology's Innovation in Science Pursuit for Inspired Research (DST-INSPIRE) program in India. M.E.'s work is supported by JST EXPERT-J, Japan Grant Number JPMJEX2510, P.M.'s work is supported by the Ministry of Education under the Rashtriya Uchchatar Shiksha Abhiyan (MoE RUSA 2.0) program at Bharathidasan University, specifically in the Physical Sciences.

	\appendix

	\section{Numerical technique to solve the hydrodynamic equation.}
	In this Appendix, we present the numerical method employed to solve the hydrodynamic equation governing collective excitations. The approach is based on reformulating the problem as a quadratic eigenvalue problem, which allows for an efficient calculation of the excitation frequencies of the system. 
	We begin by expressing the original hydrodynamic equation Eq.~\ref{eq:hydro_eig} as a quadratic eigenvalue problem of the following form
	\begin{align}
		\label{eq:QEP}
		\left( \lambda^{2} M_{2} + \lambda M_{1} + M_{0} \right) f = 0,
	\end{align}
	where the matrices are defined as
	\begin{align}
		M_{2} &= I, \\
		M_{1} &= 2\,\mathrm{diag}[B(r)], \\
		M_{0} &= L + \mathrm{diag}\!\left[
		B(r)^{2}
		- \left(g - C\nu\right)\frac{l^{2} n_{0}(r)}{r^{2}}
		\right],
	\end{align}
	and
	\begin{align}
		B(r)
		=
		l\left(\Omega + C\,n_{0}(r)\right) 
		- \frac{l\nu}{r^{2}} .
	\end{align}
	
	The linearized hydrodynamic operator in $M_{0}$ has the form
	\begin{align}
		\mathcal{L} f(r)
		&=
		\left(g - C\nu\right)\frac{1}{r}
		\frac{d}{dr}
		\!\left[
		p(r)\frac{df}{dr}
		\right],\\ \nonumber
		p(r)&=n_{0}(r)\, r.
	\end{align}
To discretize the operator, we employ a conservative finite-difference scheme based on midpoint fluxes. In this scheme, the flux at the midpoint between grid points $r_i$ and $r_{i+1}$ is denoted as:
	\begin{align}
		p_{i+\frac12} = p(r_{i}+h/2),
	\end{align}
	and denote the flux evaluated at the midpoint between $r_{i}$ and $r_{i+1}$, with $h = (R_{\rm out}-R_{\rm in})/(N+1)$.
	The discrete operator is then given by
	\begin{align}
		(\mathcal{L} f)_{i}
		=
		\frac{1}{h^{2} r_{i}}
		\left[
		p_{i+\frac12}(f_{i+1}-f_{i})
		-
		p_{i-\frac12}(f_{i}-f_{i-1})
		\right].
	\end{align}
This yields a symmetric tridiagonal matrix $L$ with elements
	\begin{align}
		L_{i,i-1}
		&= \frac{p_{i-\frac12}}{h^{2} r_{i}}, \\
		L_{i,i}
		&= -\frac{p_{i+\frac12}+p_{i-\frac12}}{h^{2} r_{i}}, \\
		L_{i,i+1}
		&= \frac{p_{i+\frac12}}{h^{2} r_{i}}.
	\end{align}
To handle the boundaries, we impose vanishing-flux (Neumann) boundary conditions at the inner and outer radii, $R_{\rm in}$ and $R_{\rm out}$. This ensures that there is no flux through the boundary, a common condition in many physical systems, especially in confined geometries. The frequencies of the collective modes are then obtained by solving the resulting finite-dimensional generalized eigenvalue problem.

\bibliography{reference}
\end{document}